\documentclass[twocolumn,floatfix,prd,authoryear,nofootinbib,aps,10pt,tightlines]{revtex4-2}

\usepackage{bm}
\usepackage{newtxmath,newtxtext}
\usepackage{graphicx}
\usepackage{textcomp}
\usepackage{float}
\usepackage{booktabs}
\usepackage{dcolumn,enumerate}
\usepackage{ragged2e}
\usepackage{hyperref}
\hypersetup{colorlinks,citecolor=blue}
\hypersetup{colorlinks=true, linkcolor=blue, filecolor=blue, urlcolor=blue}

\usepackage{xcolor}
\usepackage{epsfig}
\usepackage{caption}
\usepackage{appendix}
\usepackage{subcaption}
\usepackage{graphicx} 

\begin{document}

\title{Constraints on extra charges in dyonic Kerr-Newman-Kasuya-Taub-NUT black hole from the observations of quasi-periodic oscillations}

\author{Hamza Rehman ${}^{a, b}$}
\email{hamzarehman244@zjut.edu.cn}
\author{Saddam Hussain ${}^{a, b}$}
\email{saddamh@zjut.edu.cn}
\author{Ghulam Abbas ${}^{c}$}
\email{ ghulamabbas@iub.edu.pk}
\author{Tao Zhu ${}^{a, b}$}
\email{Corresponding author: zhut05@zjut.edu.cn}

\affiliation{${}^{a}$Institute for Theoretical Physics and Cosmology, Zhejiang University of Technology, Hangzhou 310023, China}
\affiliation{${}^{b}$United Center for Gravitational Wave Physics (UCGWP), Zhejiang University of Technology, Hangzhou, 310023, China}
\affiliation{${}^{c}$Department of Mathematics, The Islamia University of Bahawalpur, Bahawalpur, Pakistan}

\date{\today}

\begin{abstract}

This paper investigates the influence of the dimensionless electric charge ($Q/M$), magnetic charge ($P/M$), and Taub–NUT parameter ($n/M$) of a dyonic Kerr–Newman–Kasuya–Taub–NUT black hole on quasi-periodic oscillations (QPOs) observed in X-ray binaries. Using the relativistic precession model, we calculate the three fundamental frequencies arising from particle motion in the accretion disk around the black hole. These theoretical predictions are then confronted with observational QPO data from five X-ray binaries (GRO J1655–40, XTE J1859+226, XTE J1550–564, GRS 1915+105, and H1743–322), and the Markov Chain Monte Carlo technique is used to constrain the black hole parameters. Our analysis reveals no significant evidence for nonzero values of $Q/M$ and $P/M$ across all sources, thereby allowing us to place several stringent upper limits on electric charge ($Q/M$) and magnetic charge ($P/M$) of the black hole in these systems. Similarly, no compelling indication of a nonzero Taub–NUT parameter is found in QPOs from GRO J1655–40, XTE J1859+226, XTE J1550–564, and H1743–322. In contrast, the posterior distribution derived from GRS 1915+105 data suggests a nonzero Taub–NUT parameter, i.e., gravitomagnetic monopole moment. This result indicates a potential deviation from the Kerr metric in this astrophysical black hole.

\end{abstract}

\maketitle
\section{Introduction}
\renewcommand{\theequation}{1.\arabic{equation}} \setcounter{equation}{0}

Black holes (BHs) are one of the most fascinating and intriguing topics in modern physics, due to their remarkable characteristics and the profound questions they raise about the nature of gravity and spacetime. A significant breakthrough in this field was achieved in 2019, when the Event Horizon Telescope (EHT) collaboration successfully captured the first image of the supermassive BH at the center of the M87$^{*}$ galaxy \cite{EventHorizonTelescope:2019dse, EventHorizonTelescope:2019uob}. This achievement marked a major milestone, enabling researchers to directly test theoretical predictions of BH shadows against observational evidence \cite{Gan:2021xdl}. Building on this success, the EHT presented the first image of Sgr A$^{*}$—the supermassive BH at the center of our own Milky Way—in 2022, further deepening our understanding of these enigmatic objects \cite{Lu:2018uiv, EventHorizonTelescope:2022wok, Grigorian:2024rsn}. The EHT is utilized to observe supermassive BHs located in the centers of galaxies. To accomplish this, scientists combine data from numerous radio telescopes located around the globe. Collectively, these telescopes form a virtual instrument with a resolution equivalent to the Earth's diameter. This unprecedented angular resolution allows for the direct observation of structures at scales comparable to a BH’s event horizon. 

In addition to direct imaging, quasi-periodic oscillations (QPOs) observed in X-ray binaries provide a powerful tool for probing the spacetime geometry around the BH and the nature of gravity in the strong field regime. These QPOs, first identified as a distinct astronomical phenomenon in the 1980s \cite{1979Natur.278..434S}, arise from the relativistic motion of matter within the accretion disk spiraling inwards toward the BH. High-precision X-ray timing observations of these oscillations in BH X-ray binaries enable detailed investigations of the spacetime geometry and strong-field gravity in the vicinity of compact objects \cite{Stella:1998mq, Stella:1997tc}. Typically, an X-ray binary system comprises a compact object—either a BH or a neutron star—and a companion donor star. Matter transferred from the companion accretes onto the compact object, forming a hot accretion disk that emits X-rays. The presence of QPOs, manifesting as periodic modulations in X-ray emission, provides invaluable insight into the dynamics of the accretion flow and the extreme gravitational environment near compact objects. Remarkably, the timing sensitivity of these observations probes spatial scales down to sub-nano-arcseconds, far surpassing what is achievable with current imaging techniques \cite{Ingram:2019mna, Remillard:2006fc}. 

There are several theoretical models of the physical mechanism responsible for the production of QPOs, like the relativistic precession models \cite{Stella:1997tc, Stella:1998mq, Stella:1999sj}, the resonance models \cite{Abramowicz:2003xy, Abramowicz:2001bi, Kluzniak:2001ar}, and the diskoseismology models \cite{Perez:1996ti, Silbergleit:2000ck, Wagoner:2001uj}. Here we focus our attention on the relativistic precession model that fits qualitatively well in the widespread data given by observations of a variety of BH and neutron star binaries. The relativistic precession model \cite{Stella:1998mq, Stella:1997tc, Stella:1999sj} is related to the three frequencies that describe the motion of test particles in the vicinity of the compact object in a strong gravitational field. This model is a well-accepted theoretical framework that is utilized to explain the X-ray QPO phenomena. The frequencies include the orbital, vertical, and radial epicyclic frequencies, which describe the motion of a test particle in the azimuthal direction, oscillations perpendicular to the orbital plane, and oscillations toward and away from the central object. 

To provide information regarding strong-field gravitational effects, X-ray signals are emitted from the gas particles that are accreting around a central object at distances of numerous or tens of gravitational radii. These X-ray signals are emitted from gas particles spiraling around the central object, typically located at radii of several to tens of gravitational radii from the compact object. The detected QPOs contain useful information regarding relativistic effects in the strong field. The original purpose of this model was to explain the high-frequency QPOs observed in X-ray binaries containing neutron stars; subsequently, it has been extended for stellar-mass BHs \cite{Stella:1998mq}.

The observation of QPOs in BH binaries is comparatively rarer than in neutron star systems. Despite this, BH offers a relatively clean astrophysical environment that provides an excellent laboratory for studying a strong gravitational field and the spacetime geometry in the vicinity of the BH \cite{Motta:2013wga}. Several studies have revealed that the relativistic precession model effectively explains the observed data for three different frequencies in BH X-ray binaries and is used for measuring BH mass and spin, as found in numerous sources \cite{Motta:2013wga, Motta:2022rku, Ingram:2014ara, Remillard:2002cy}. Several studies have delved into the theoretical aspects of this subject. Some examples include testing the no-hair theorem with GRO J1655-40 \cite{Allahyari:2021bsq}, analyzing a BH in non-linear electrodynamics \cite{Banerjee:2022chn}, investigating the BH candidate \cite{Bambi:2012pa, Bambi:2013fea}, examining the behavior of QPOs around rotating wormholes \cite{Deligianni:2021ecz, Deligianni:2021hwt}, and conducting tests of gravity within different modified gravity theories \cite{Maselli:2014fca, Chen:2021jgj, Wang:2021gtd, Jiang:2021ajk, Ashraf:2025lxs, Mustafa:2025jco, Yang:2025aro, Guo:2025zca, Yang:2024mro, Liu:2023ggz, DeFalco:2023kqy, Bambi:2022dtw, Liu:2023vfh}.

This study investigates QPOs in the spacetime geometry of the dyonic Kerr-Newman-Kasuya-Taub-NUT (KNKTN) BH \cite{Ali:2007xz}. Compared to the standard Kerr BH, the dyonic KNKTN solution incorporates several additional parameters, including the NUT charge, electric charge, and magnetic charge, in addition to the mass and spin. Although QPO signals observed in many sources can be well explained by the relativistic precession model applied to Kerr BHs, these same signals offer a powerful tool to constrain the presence of extra charges predicted in the dyonic KNKTN metric. Our aim here is to explore how the interplay of these extra parameters impacts the observable characteristics of QPOs in X-ray emissions from accreting systems. This is achieved by analyzing geodesic motion, epicyclic frequencies, and their observational constraints.

The concept of dyonic BHs has gained attention because of developments in quantum field theory and classical general relativity. Several decades ago, Dirac \cite{Dirac:1931kp, Dirac:1948um} proposed the idea of magnetic monopoles, which are crucial to understanding dyonic configurations. Subsequently, the concept of magnetic monopoles has been shown to arise naturally in grand unified theories (GUTs) as topological solutions in non-Abelian gauge fields \cite{tHooft:1974kcl, Polyakov:1974ek}. These theoretical advances established the foundation for the incorporation of magnetic charges into BH solutions and explored their implications in astrophysical and high-energy contexts.
The Dyonic BH solutions in general relativity were initially proposed by Carter \cite{4a}, who introduced a magnetic charge parameter to the Reissner-Nordstrom (RN) and Kerr-Newman \cite{Newman:1965my} spacetimes for physical grounds.

Furthermore, the Taub-NUT metric is commonly recognized as the gravitational counterpart of an electromagnetic dyon \cite{Lynden-Bell:1996dpw, Nouri-Zonoz:1997mnd, Nouri-Zonoz:1998whb}, as Ashtekar-Magnon-Das (AMD) charges represent the major contributions of the electric and magnetic components of the Weyl tensor \cite{Ashtekar:1984zz, Ashtekar:1999jx}. 
The dyonic KNTN spacetime is an electrovacuum solution of the Plebanski-Demianski family \cite{Plebanski:1976gy}. The Killing-Yano tensor leads to hidden symmetry in these spacetimes \cite{Demianski:1980mgt}. The higher-dimensional generalization of the KNTN-AdS BH has been discussed \cite{Chen:2006xh}, while \cite{Kubiznak:2006kt} explored its hidden symmetries. Furthermore, the analysis of the integrability of geodesic motion in arbitrary dimensions has been addressed in \cite{Page:2006ka}, and the separability of Hamilton-Jacobi and Klein-Gordon equations on this background has been proven in \cite{Dadhich:2001sz, Frolov:2006pe}. In the absence of Maxwell fields, the thermodynamics of this solution, along with a generalized Smarr relation and its Brown-York charges, are derived in Ref. \cite{Rodriguez:2021hks}. The supersymmetric properties of this solution have been investigated in Ref.~\cite{Alonso-Alberca:2000zeh}, demonstrating that they typically preserve 1/4 of the total supersymmetry.

The structure of this paper is as follows: In Section II, we present a basic derivation of the QPO frequencies from the Euler–Lagrange equation of motion for massive particles in the dyonic KNKTN spacetime. Section III covers the X-ray QPO observation data, Markov chain Monte Carlo (MCMC) analysis, and the best-fit values obtained from the MCMC simulation to constrain the parameters of the BH. Finally, in Section IV, we provide the conclusions of our work.

\section{QPOS FREQUENCIES IN THE DYONIC KNKTN BH}
\renewcommand{\theequation}{2.\arabic{equation}} \setcounter{equation}{0}

The line element of the dyonic KNKTN BH in Boyer-Lindquist coordinates can be expressed as \cite{Ali:2007xz}
\begin{eqnarray}
ds^2 &=& -\frac{\Delta_r}{\Sigma}(dt - \eta d\varphi)^2 + \frac{\Sigma}{\Delta_r}dr^2 + \Sigma d\theta^2 \nonumber\\
&&+ \frac{\sin^2\theta}{\Sigma}(a dt - \rho^2 d\varphi)^2, \label{za1}
\end{eqnarray}
where
\begin{eqnarray}
\Sigma &=& r^2 + (n + a \cos\theta)^2,\\
\eta &=& a \sin^2\theta - 2n \cos\theta,\\
\Delta_r &=& \rho^2 - 2(Mr + n^2) + Q^2 + P^2,\\
\rho^2 &=& r^2 + a^2 + n^2.
\end{eqnarray}
Here, $M$ denotes the mass, $a= J/M$ represents angular momentum per unit mass with $J$ being the angular momentum, and $Q$, $P$, and $n$ represent three extra charges of the dyonic KNKTN BH, namely the electric charge, magnetic charge, and Taub-NUT parameter, respectively.

One can compute the orbital, periastron precession, and nodal precession frequencies to explain QPOs by applying the relativistic precession model and utilizing the equation of motion of a test particle in the accretion around the dyonic KNKTN BH. Accretion disks form when matter accumulates in nearly circular orbits around a compact object, and their characteristics are determined by the geometry of the surrounding space-time. To investigate the fundamental frequencies that characterize QPOs, we consider the dynamics of a massive particle in the dyonic KNKTN BH background. The analysis begins with the Lagrangian that governs the dynamics of a test particle in the spacetime
\begin{equation}
\mathcal{L} = \frac{1}{2} g_{\mu\nu} \frac{dx^\mu}{d\lambda} \frac{dx^\nu}{d\lambda}, \label{za2}
\end{equation}
where $\lambda$ represents the affine parameter along the particle's worldline. For massless particle, one has $\mathcal{L}=0$, whereas for massive particle, $\mathcal{L}<0$. The associated generalized momentum is then given by
\begin{equation}
p_\mu = \frac{\partial \mathcal{L}}{\partial \dot{x}^\mu} = g_{\mu\nu} \dot{x}^\nu. \label{a1}
\end{equation}

By solving Eq.~(\ref{a1}), we obtain the equations of motion for particles with conserved energy $\tilde{E}$ and angular momentum $\tilde{L}$,
\begin{eqnarray}
p_t &=& g_{tt} \dot{t} + g_{t\phi} \dot{\phi} = -\tilde{E}, \label{a2}\\
p_\phi &=& g_{t\phi} \dot{t} + g_{\phi\phi} \dot{\phi} = \tilde{L}, \label{a3}\\
p_r &=& g_{rr} \dot{r}, \\
p_\theta &=& g_{\theta\theta} \dot{\theta}, \label{a4}
\end{eqnarray}
where a dot denotes differentiation with respect to the affine parameter $\lambda$. Solving these equations yields
\begin{eqnarray}
\dot{t} &=& \frac{g_{\phi\phi} \tilde{E} + g_{t\phi} \tilde{L}}{g_{t\phi}^2 - g_{tt} g_{\phi\phi}}, \label{a5}\\
\dot{\phi} &=& \frac{\tilde{E} g_{t\phi} + g_{tt} \tilde{L}}{g_{tt} g_{\phi\phi} - g_{t\phi}^2}. \label{a6}
\end{eqnarray}
The normalization condition for the four-velocity, $g_{\mu \nu} \, \dot{x}^{\mu} \, \dot{x}^{\nu} = -1$, leads, upon substituting Eqs.~(\ref{a5}) and (\ref{a6}), to
\begin{equation}
g_{rr} \, \dot{r}^{2} + g_{\theta\theta} \, \dot{\theta}^{2} = -1 - g_{tt} \, \dot{t}^{2} - g_{\phi\phi} \, \dot{\phi}^{2} - 2 g_{t\phi} \, \dot{t} \, \dot{\phi}. \label{a7}
\end{equation}
To simplify the analysis, we focus on equatorial motion, i.e., setting $\theta = \pi/2$ and $\dot{\theta} = 0$. Eqs.~(\ref{a5})--(\ref{a7}) then yield
\begin{equation}
\dot{r}^{2} = V_{\text{eff}}(r, M, \tilde{E}, \tilde{L}) =
\frac{\tilde{E}^{2} g_{\phi\phi} + 2 \tilde{E} \tilde{L} g_{t\phi} + \tilde{L}^{2} g_{tt}}{g_{t\phi}^{2} - g_{tt} g_{\phi\phi}} - 1, \label{a8}
\end{equation}
where $V_{\text{eff}}(r, M, \tilde{E}, \tilde{L})$ denotes the effective potential governing the radial motion of a particle with energy $\tilde{E}$ and axial angular momentum $\tilde{L}$.

Furthermore, in the equatorial plane ($\theta= \frac{\pi}{2}$), stable circular orbits are characterized by the conditions $\dot{r}=0$ and $dV_{\mathrm{eff}}/dr=0$. Imposing these requirements allows one to determine the specific energy $\tilde{E}$ and specific angular momentum $\tilde{L}$, which govern the dynamics of particles on circular orbits in the vicinity of the BH. These quantities are given by  
\begin{equation}
\tilde{E} = \frac{-g_{tt} + g_{t\phi}\Omega_\phi}{\sqrt{ -g_{tt} - 2g_{t\phi}\Omega_\phi - g_{\phi\phi}\Omega_\phi^2 }}, \label{a9}
\end{equation}
\begin{equation}
\tilde{L} = \frac{g_{t\phi} + g_{\phi\phi}\Omega_\phi}{\sqrt{ -g_{tt} - 2g_{t\phi}\Omega_\phi - g_{\phi\phi}\Omega_\phi^2 }}, \label{a10}
\end{equation}
where $\Omega_{\phi}$ denotes the angular velocity of the circular orbit, given by
\begin{equation}
\Omega_\phi = \frac{ -\partial_r g_{t\phi} \pm \sqrt{ (\partial_r g_{t\phi})^2 - (\partial_r g_{tt})(\partial_r g_{\phi\phi}) } }{ \partial_r g_{\phi\phi} }.\label{a11}
\end{equation}
Here, the upper (lower) sign corresponds to co-rotating (counter-rotating) orbits. For co-rotating orbits, the angular momentum is aligned with the spin of the BH, while for counter-rotating orbits, it is oriented opposite to the direction of spin.

\begin{figure*}
\centering
\includegraphics[width=.3\linewidth, height=1.5in]{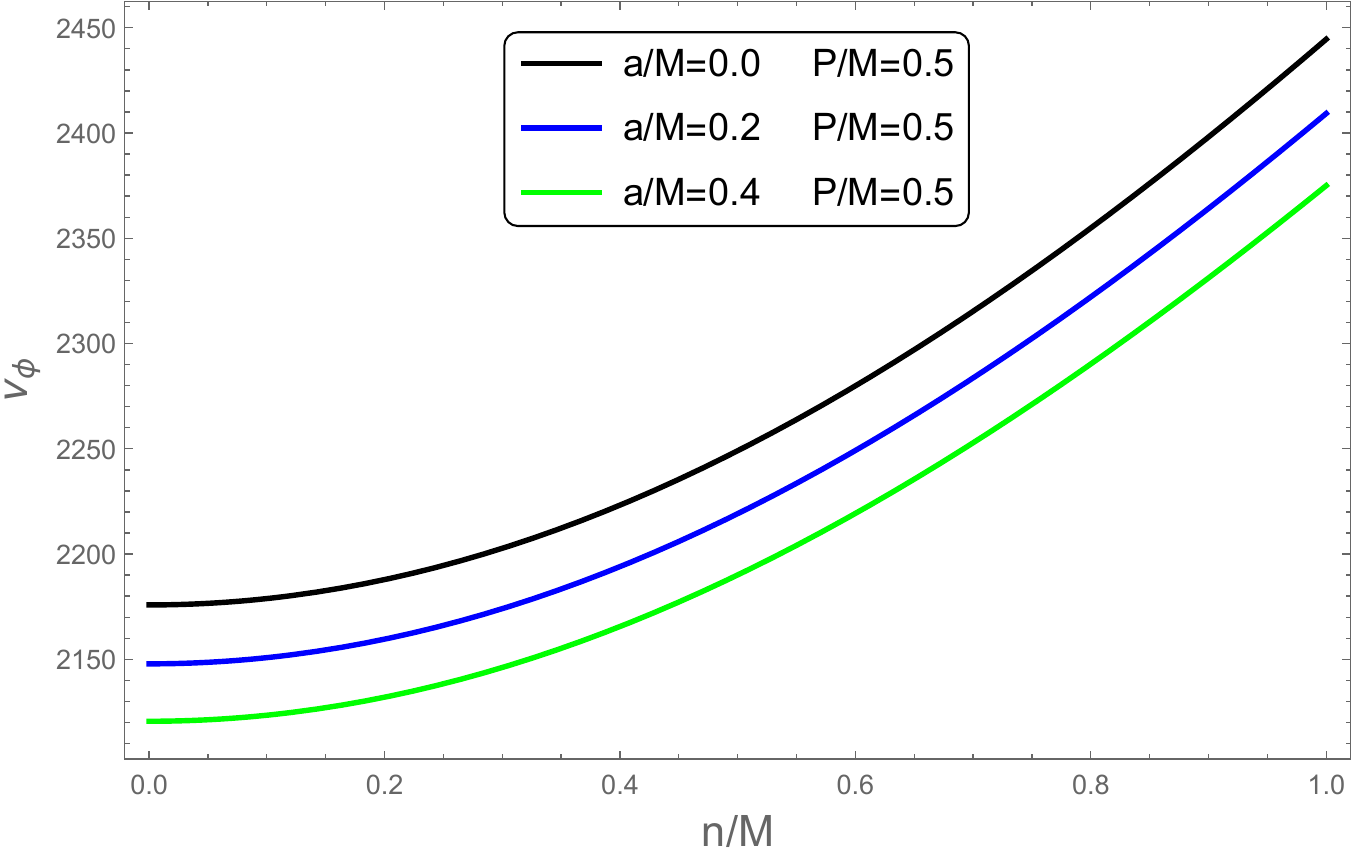}
\includegraphics[width=.3\linewidth, height=1.5in]{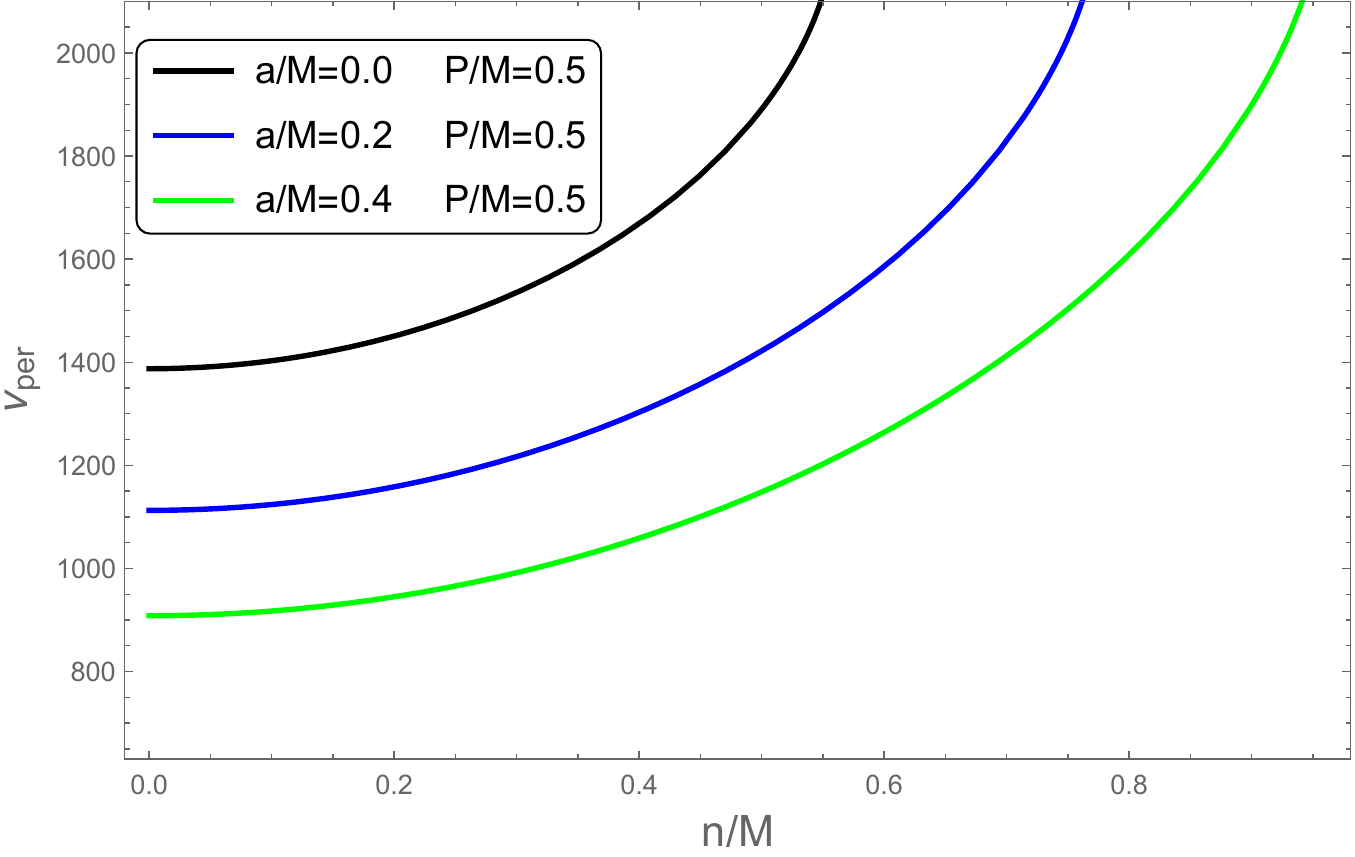}
\includegraphics[width=.3\linewidth, height=1.5in]{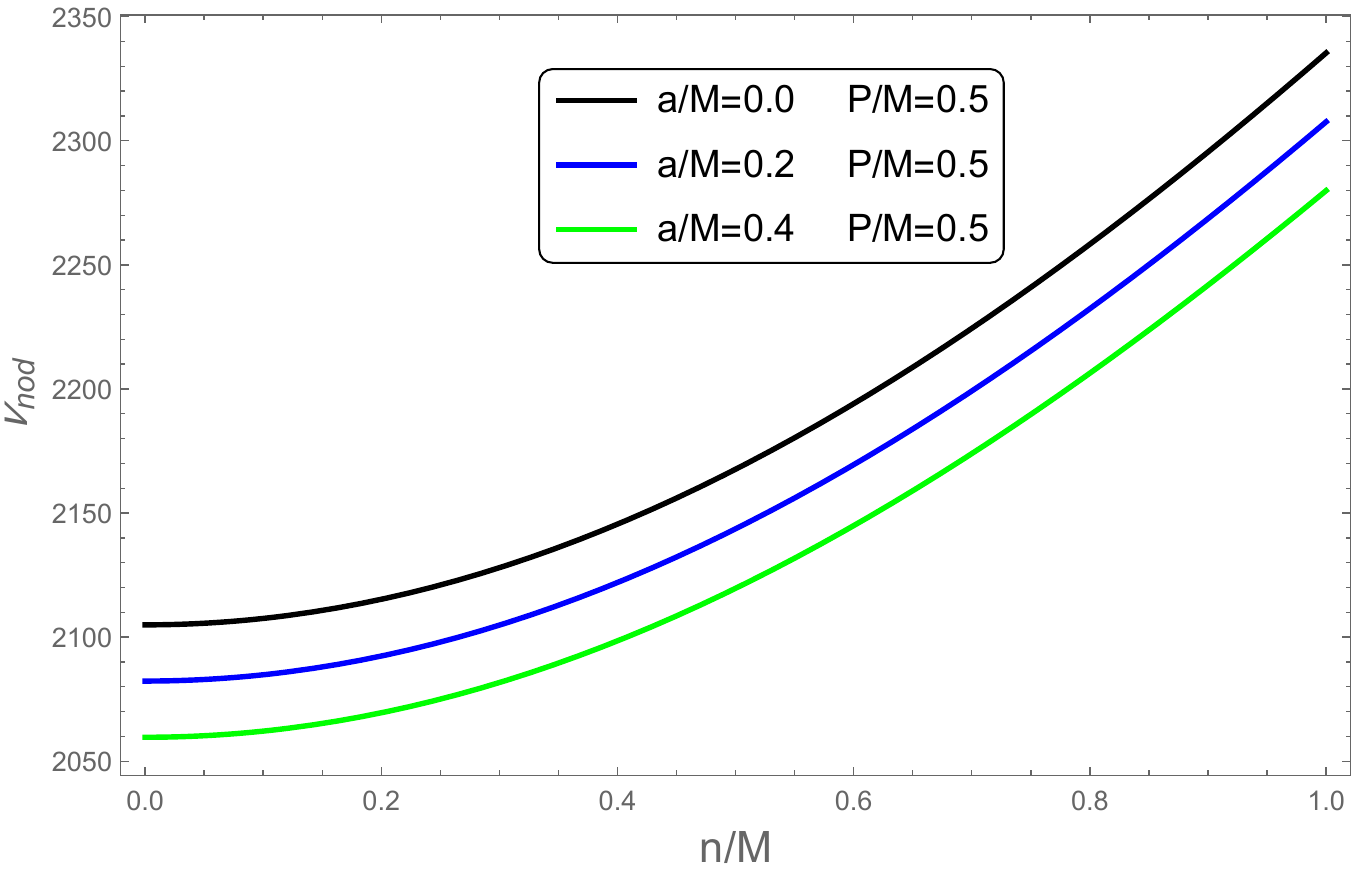}
\includegraphics[width=.3\linewidth, height=1.5in]{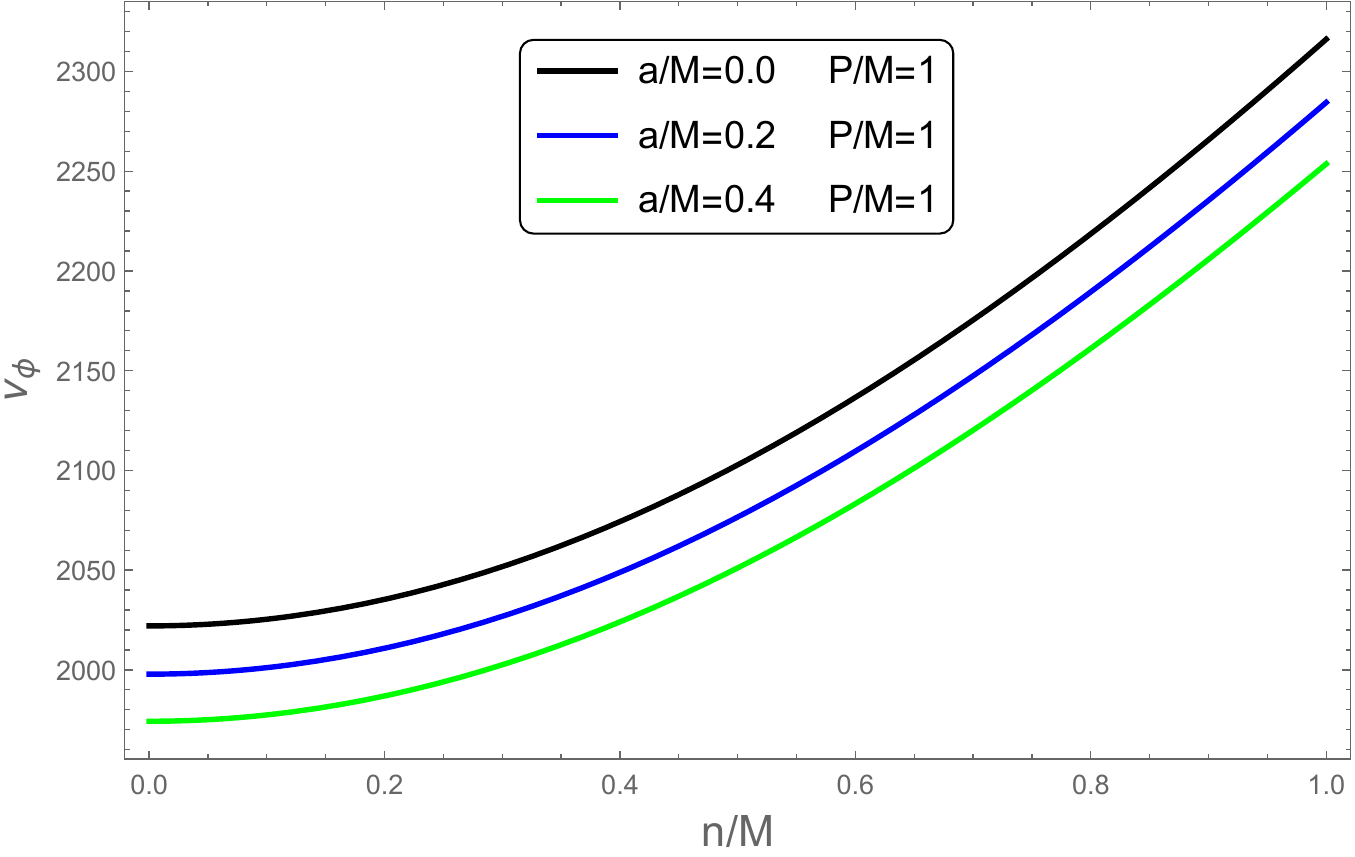}
\includegraphics[width=.3\linewidth, height=1.5in]{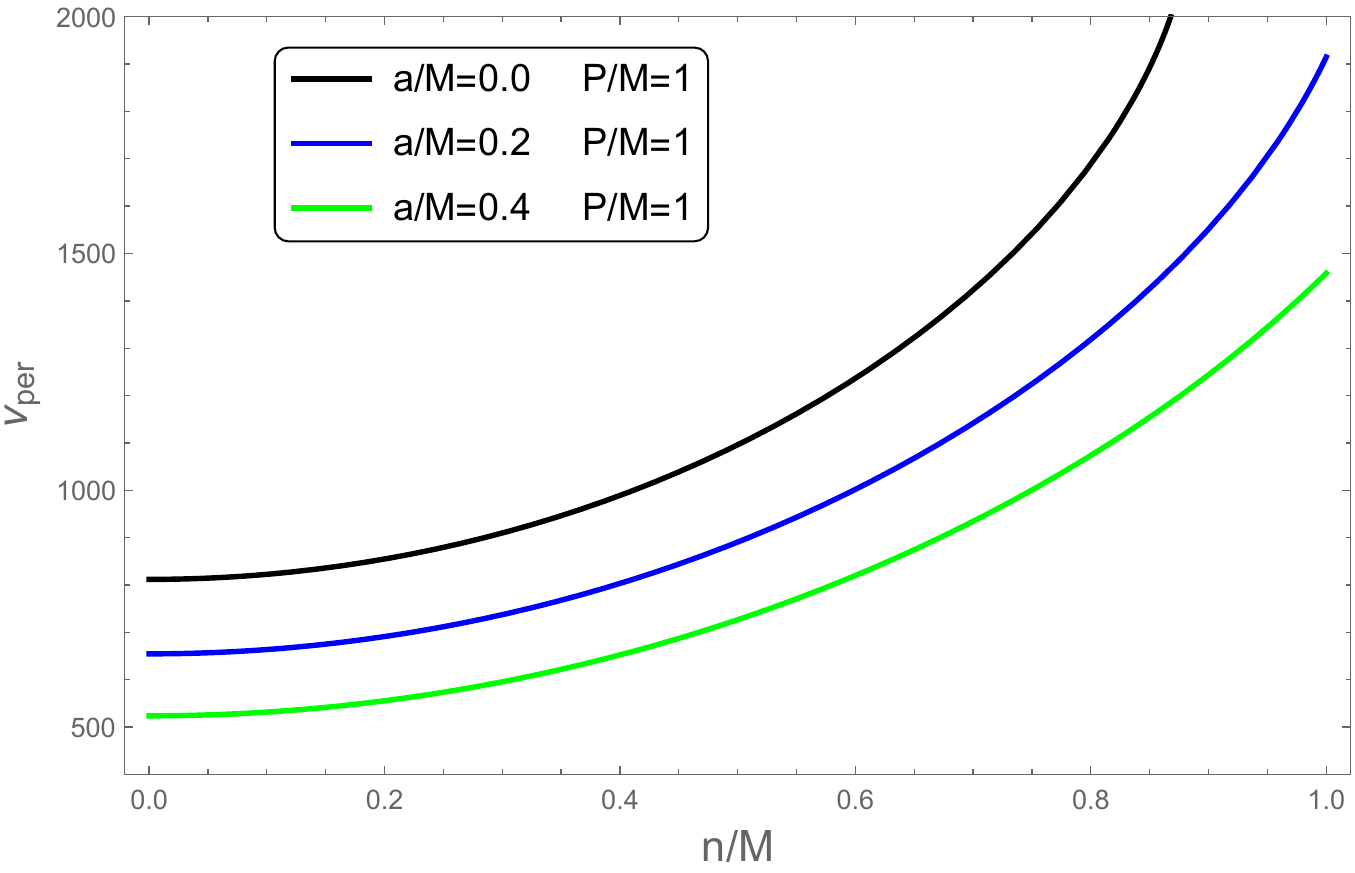}
\includegraphics[width=.3\linewidth, height=1.5in]{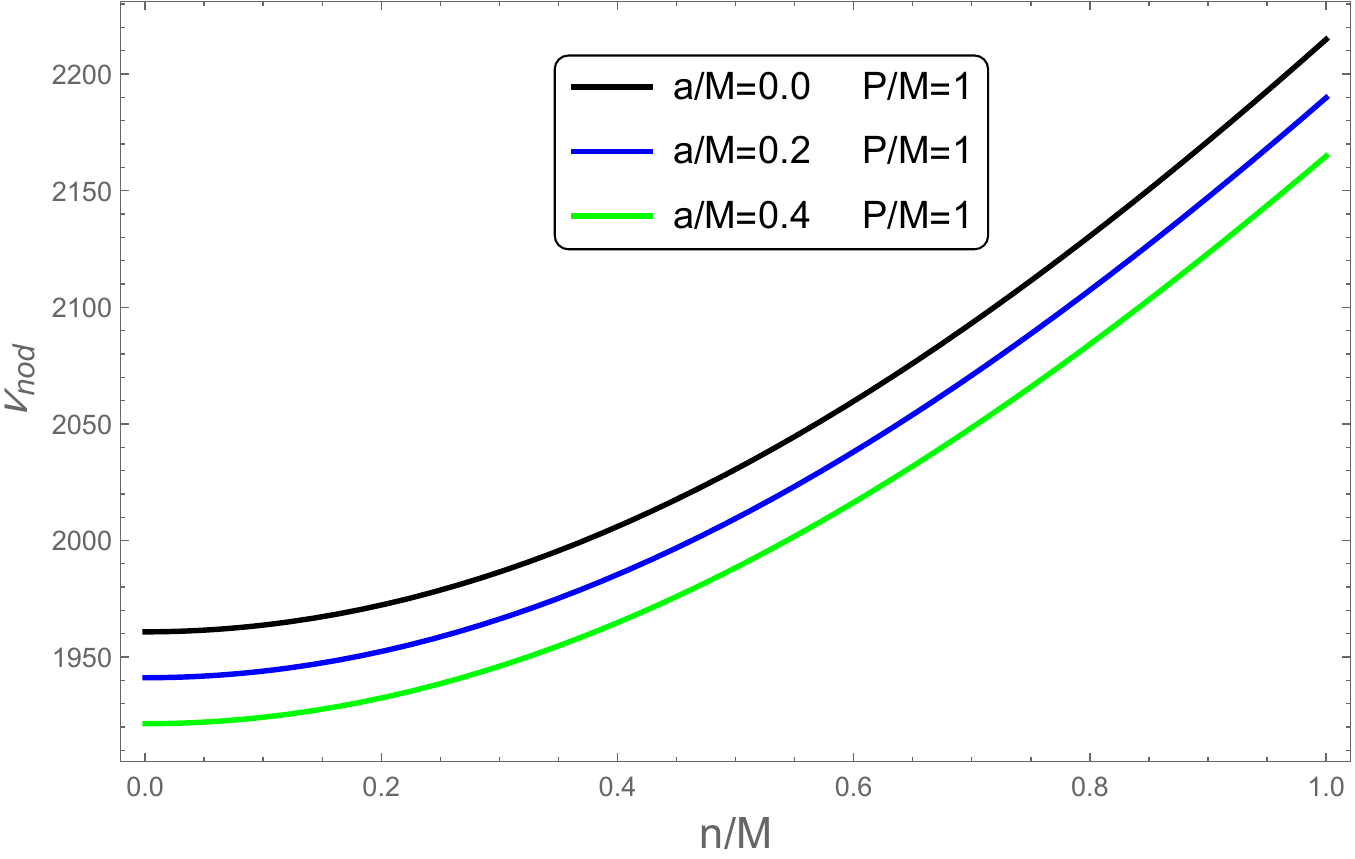}
\includegraphics[width=.3\linewidth, height=1.5in]{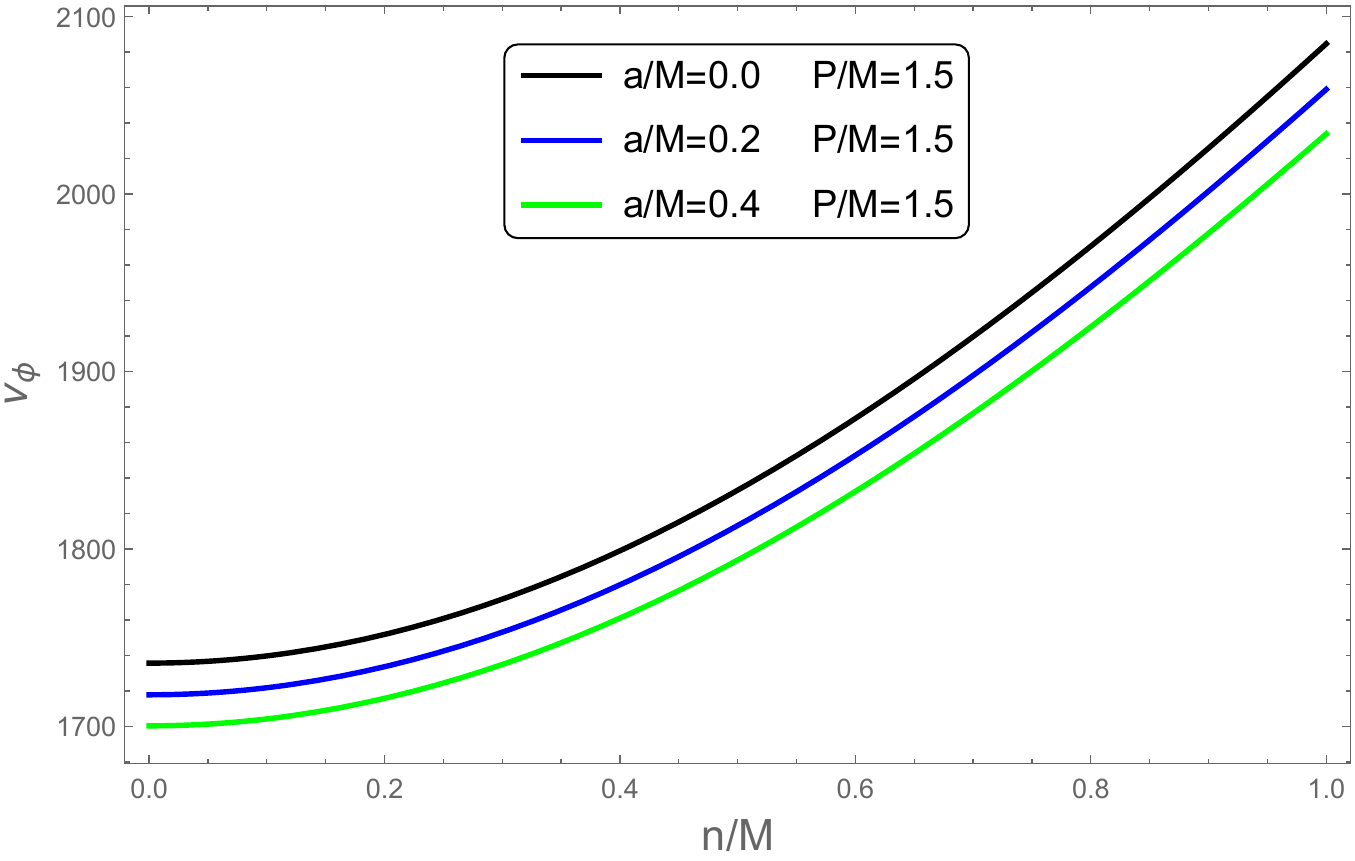}
\includegraphics[width=.3\linewidth, height=1.5in]{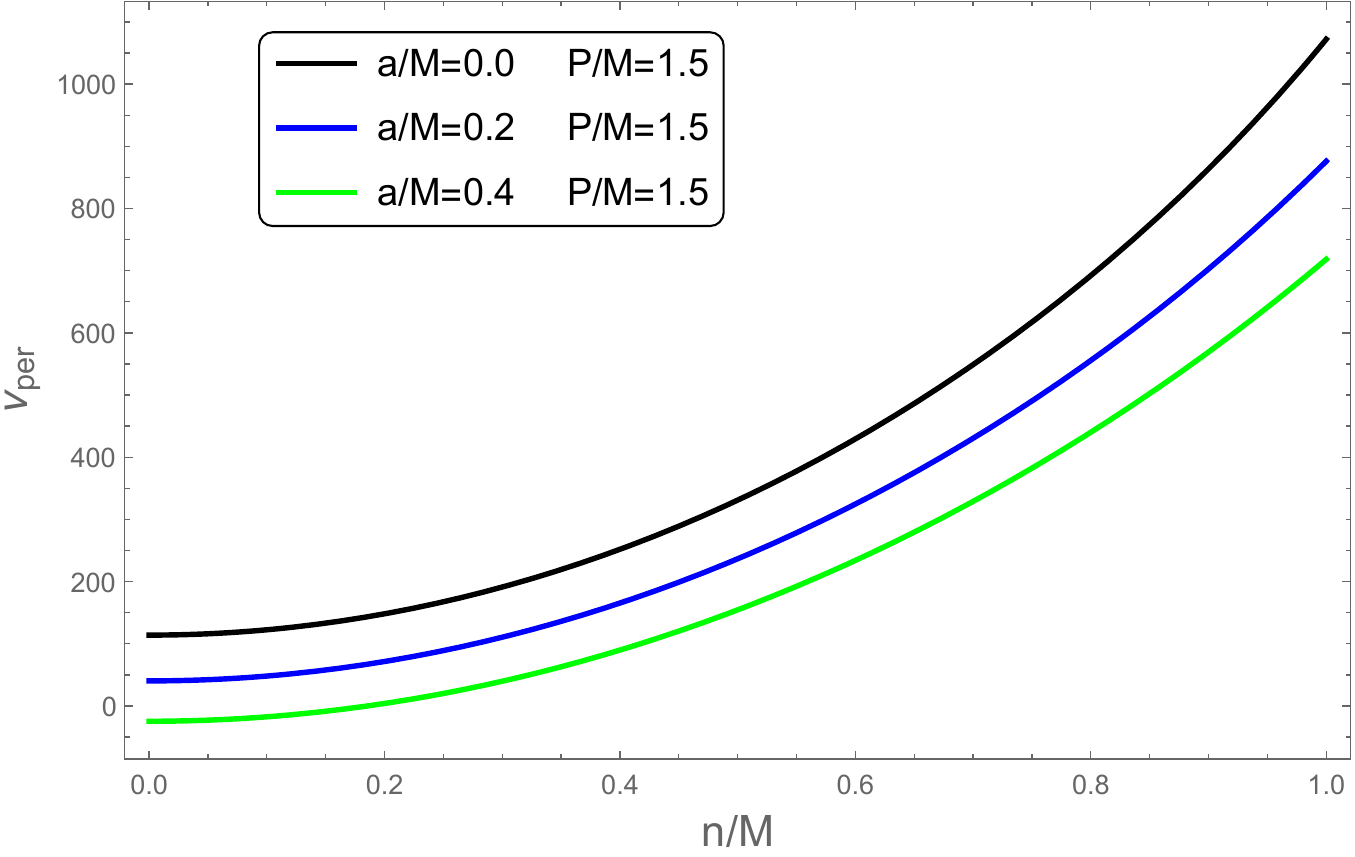}
\includegraphics[width=.3\linewidth, height=1.5in]{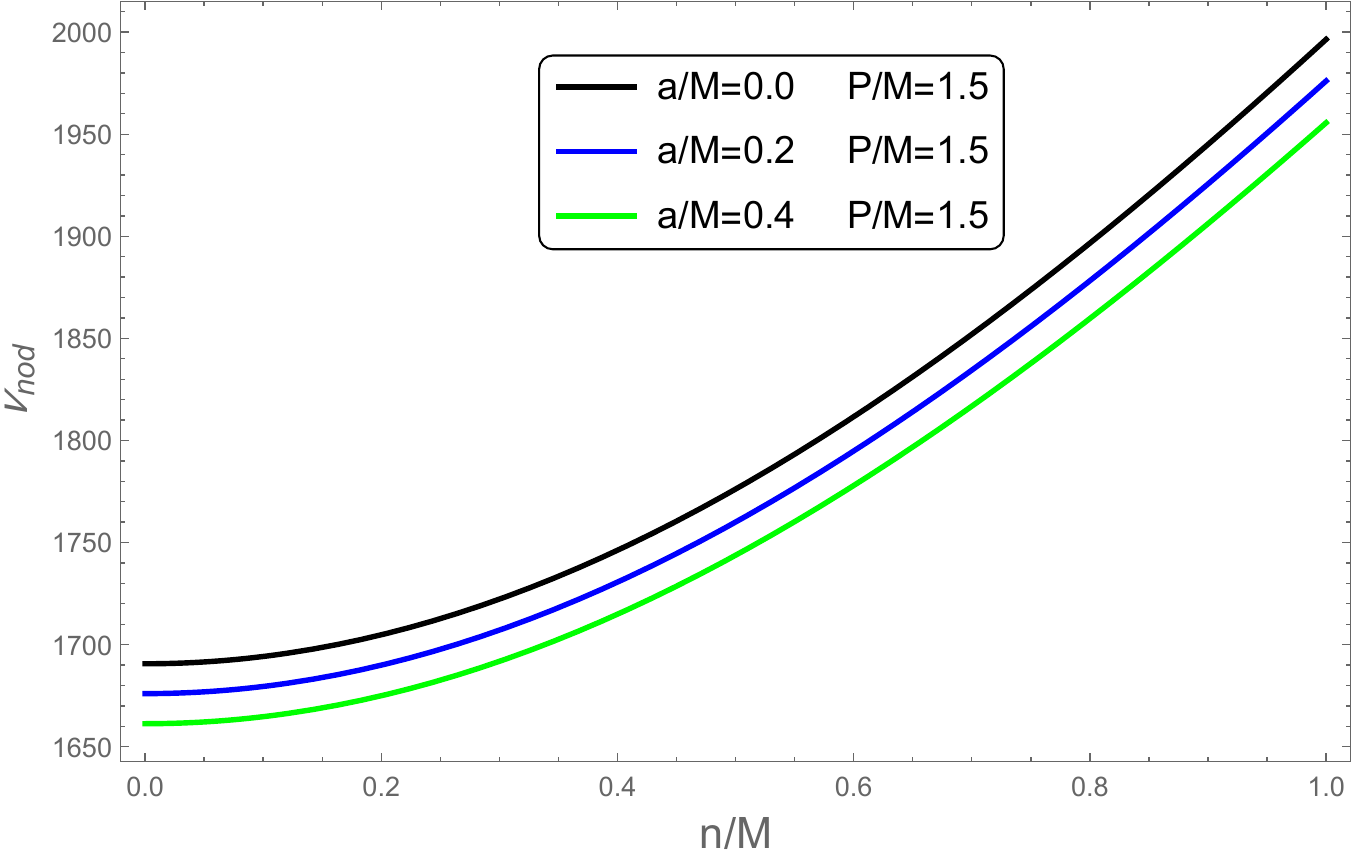}
\caption{The profile of orbital frequency $\nu_\phi$, periastron precession frequency $\nu_{\mathrm{per}}$, and the nodal precession frequency $\nu_{\mathrm{nod}}$ as a function of the dimensionless NUT parameter $n/M$ in the dyonic KNKTN BH for $P/M = 0.5$ (the top panel), $P/M = 1$ (the middle panel), and $P/M = 1.5$ (the bottom panel). By considering $M=1$, the orbit radius is $r=6.0$.
}\label{1}
\end{figure*}

\begin{figure*}
\centering
\includegraphics[width=.3\linewidth, height=1.5in]{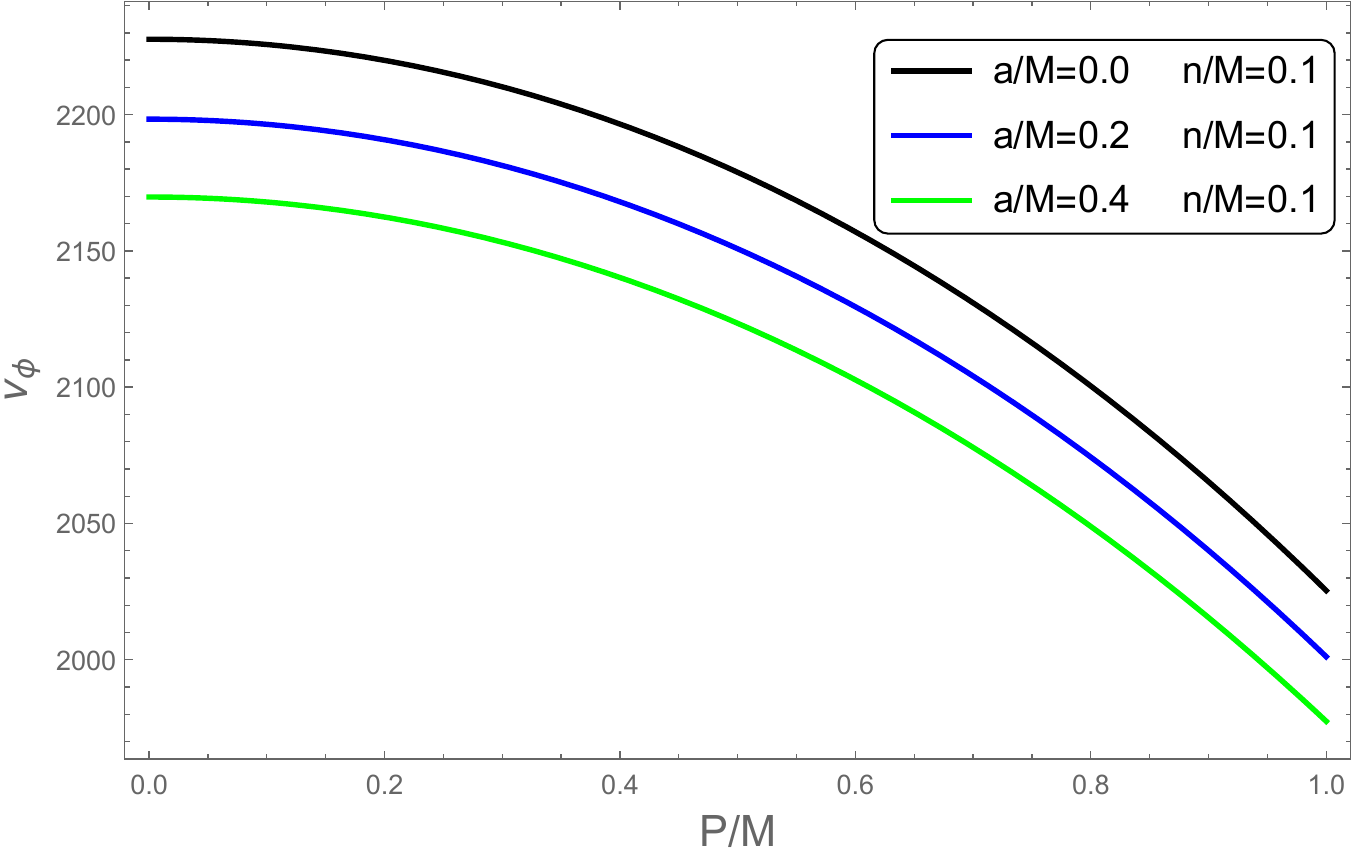}
\includegraphics[width=.3\linewidth, height=1.5in]{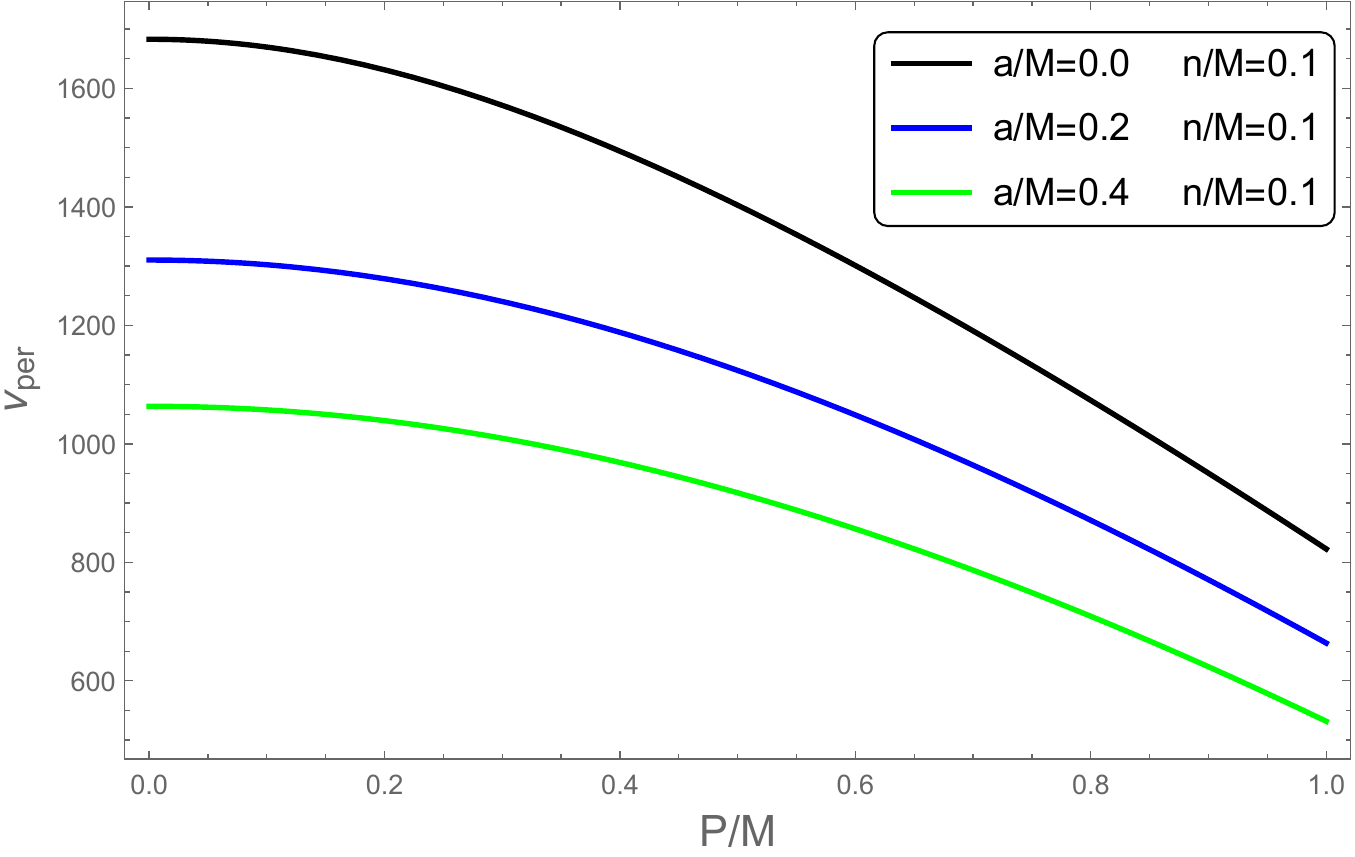}
\includegraphics[width=.3\linewidth, height=1.5in]{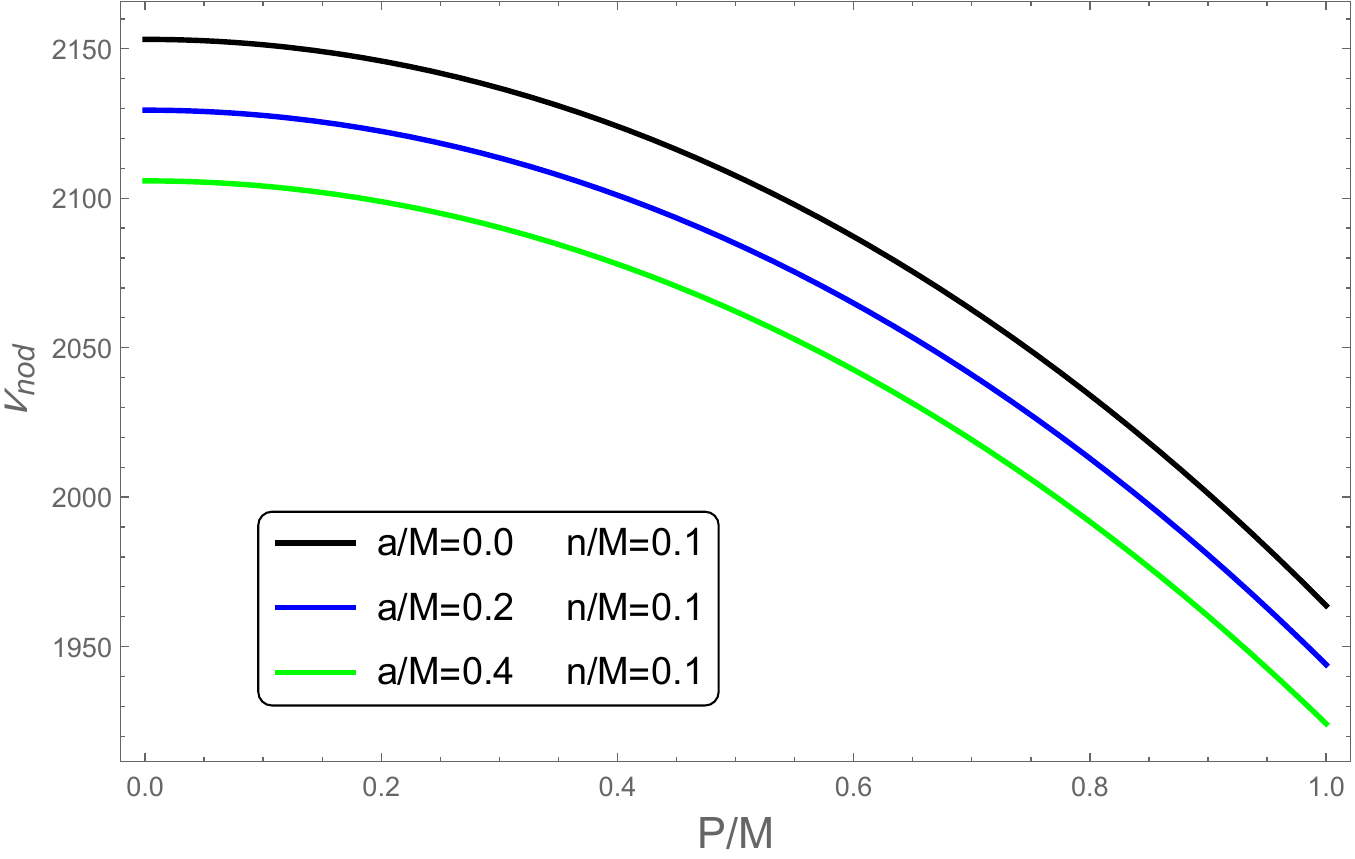}
\includegraphics[width=.3\linewidth, height=1.5in]{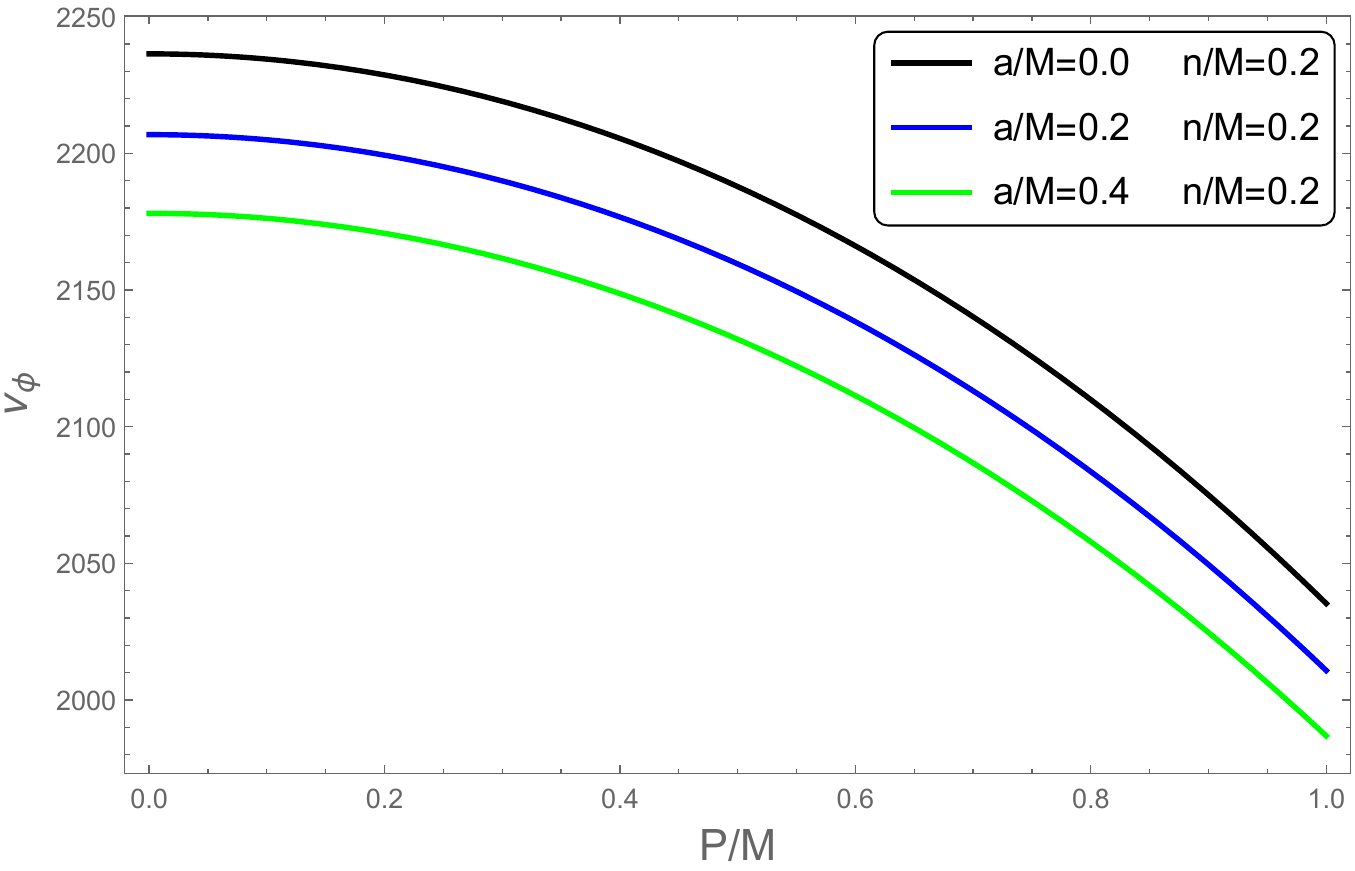}
\includegraphics[width=.3\linewidth, height=1.5in]{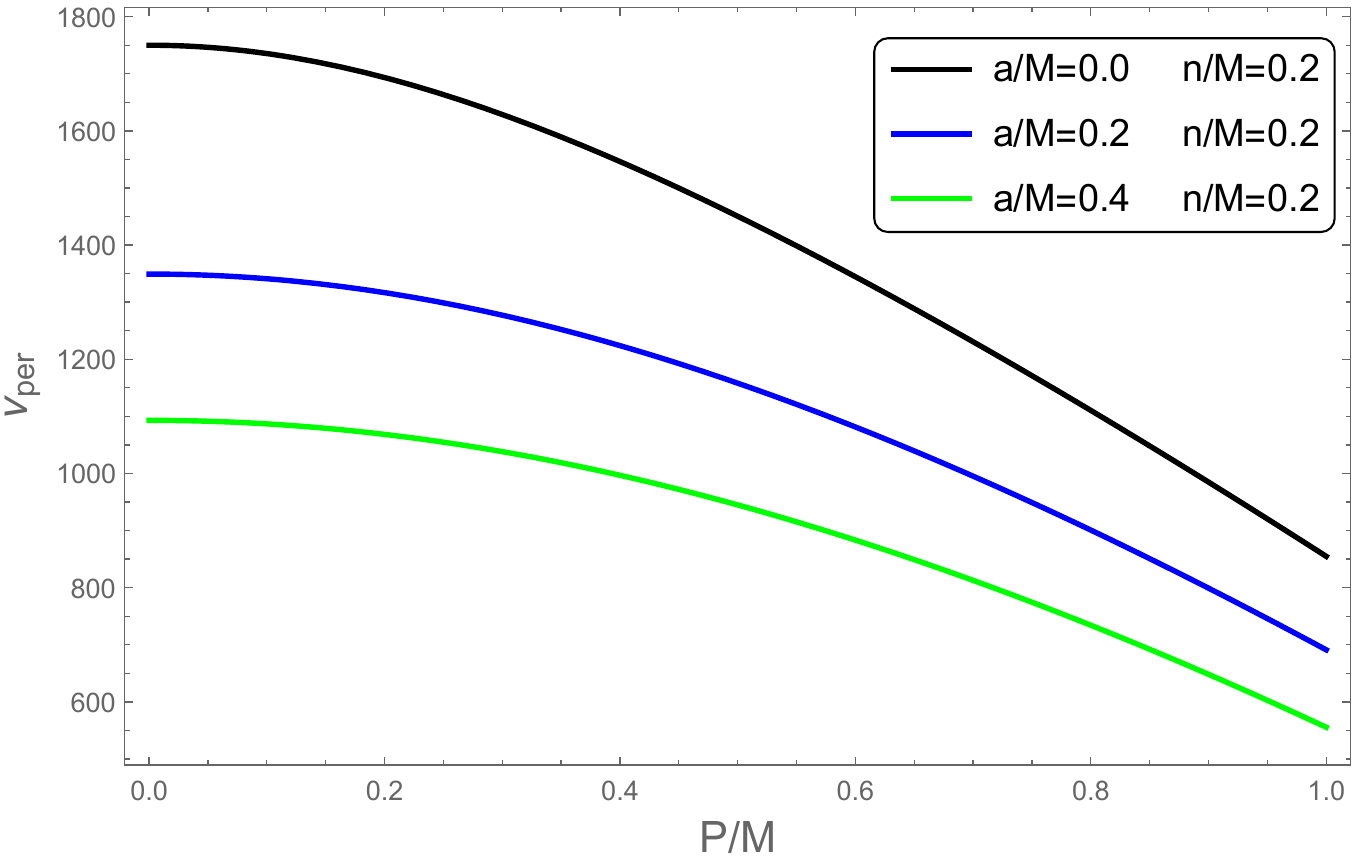}
\includegraphics[width=.3\linewidth, height=1.5in]{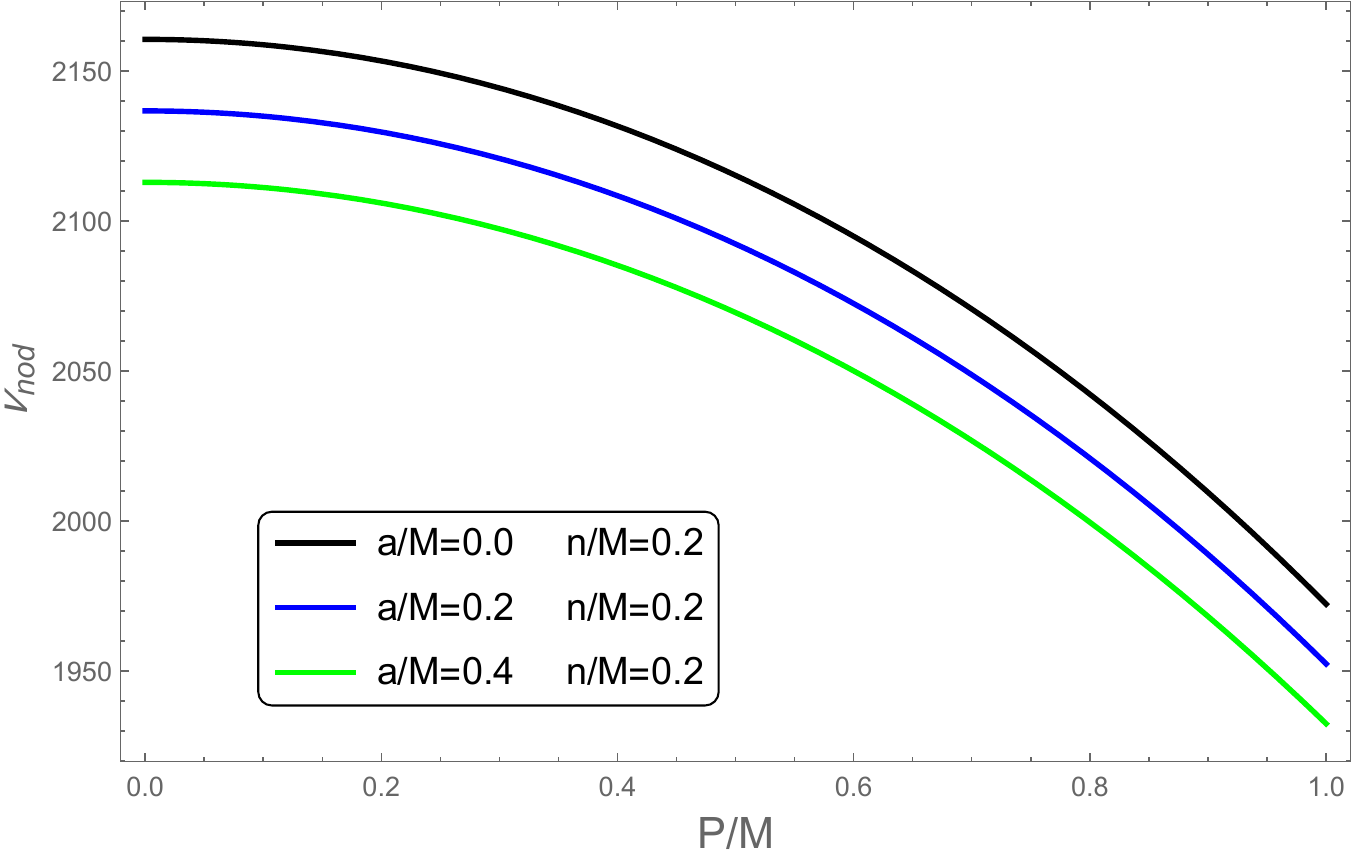}
\includegraphics[width=.3\linewidth, height=1.5in]{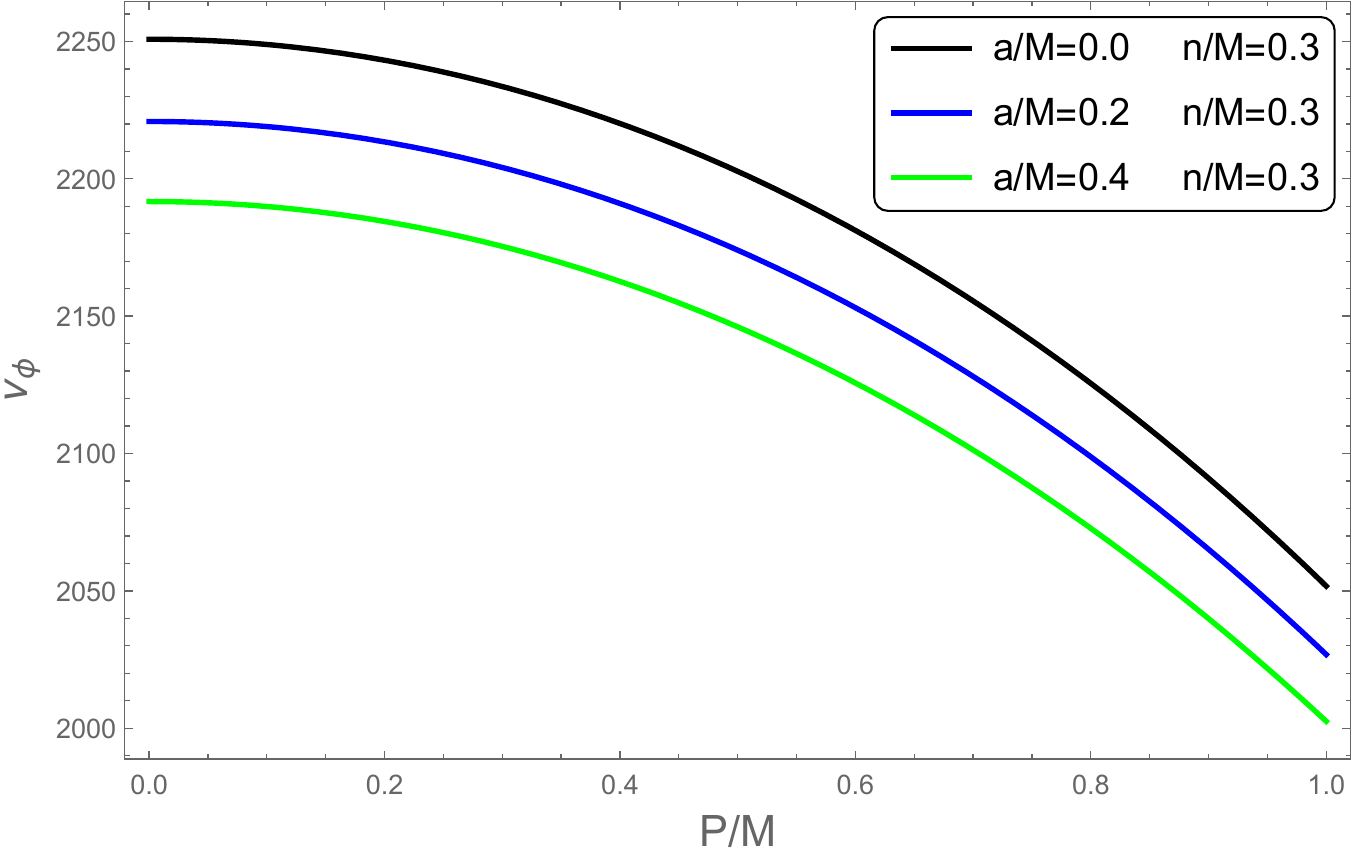}
\includegraphics[width=.3\linewidth, height=1.5in]{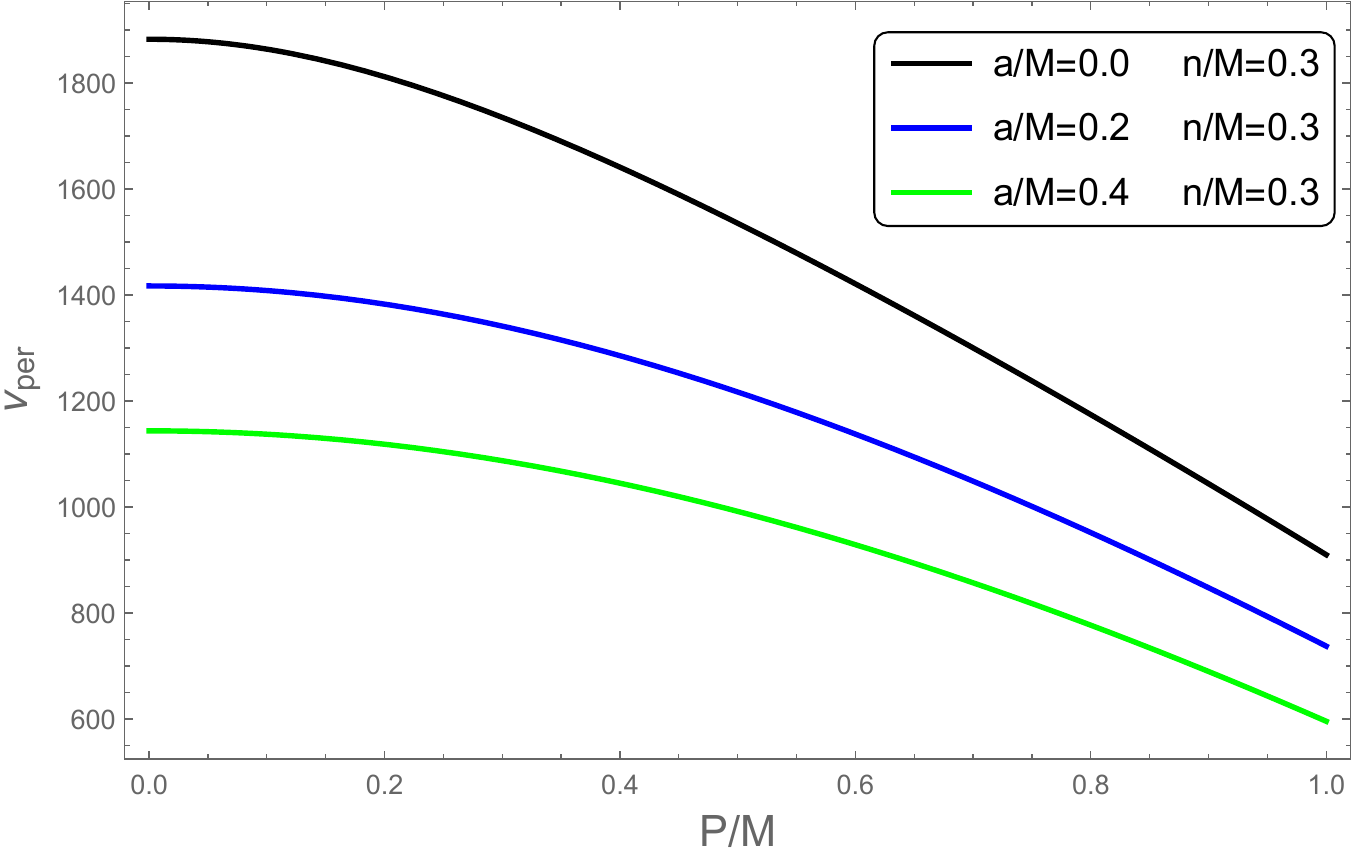}
\includegraphics[width=.3\linewidth, height=1.5in]{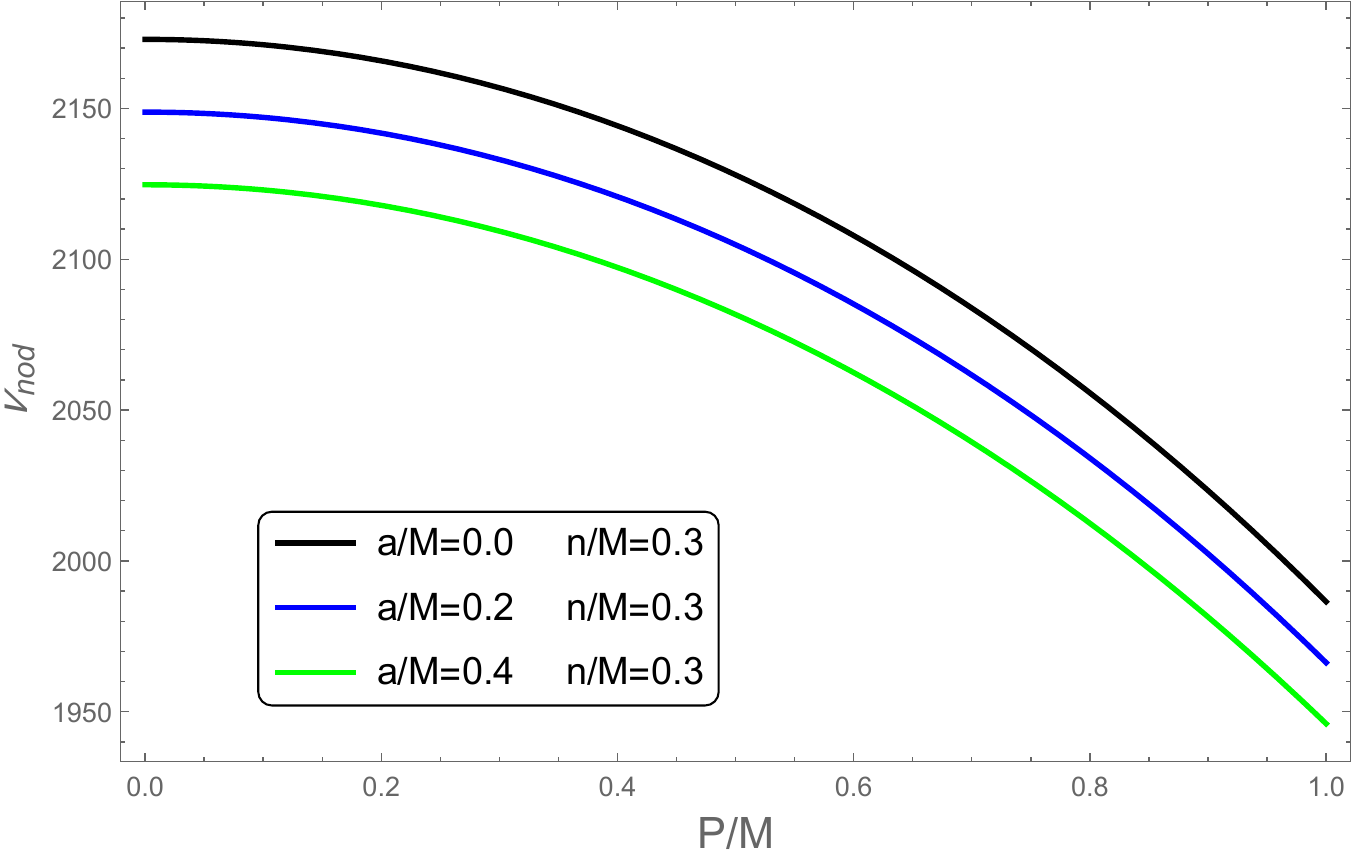}
\caption{The behavior of the orbital frequency $\nu_\phi$, periastron precession frequency $\nu_{\mathrm{per}}$, and the nodal precession frequency $\nu_{\mathrm{nod}}$ along the dimensionless magnetic charge $P/M$ in the dyonic KNKTN BH for $n/M = 0.1$ (the top panel), $n/M = 0.2$ (the middle panel), and $n/M = 0.3$ (the bottom panel). By setting $M=1$, the orbit radius is $r=6.0$.
}\label{2}
\end{figure*}

In this work, we investigate QPOs within the framework of the relativistic precession model. To this end, it is necessary to evaluate the three fundamental frequencies associated with circular equatorial orbits: the orbital frequency $\nu_\phi$, the radial epicyclic frequency $\nu_{r}$, and the vertical epicyclic frequency $\nu_{\theta}$. The orbital frequency, also referred to as the Keplerian frequency, is given by
\begin{eqnarray}
    \nu_\phi=\frac{\Omega_\phi}{2\pi}.
\end{eqnarray}
The radial and vertical epicyclic frequencies are determined by considering small perturbations about circular equatorial orbits. Introducing perturbations in the radial and polar coordinates, the motion can be expressed as
\begin{eqnarray}
r(t) = r_0 + \delta r(t), \quad \theta(t) = \frac{\pi}{2} + \delta \theta(t), \label{a14}
\end{eqnarray}
where $\delta r(t)$ and $\delta\theta(t)$ denote small deviations from the equilibrium orbit. The dynamics of these perturbations are governed by
\begin{eqnarray}
\frac{d^2 \delta r(t)}{dt^2} + \Omega_r^2 \delta r(t) = 0, \label{a15}\\
\frac{d^2 \delta \theta(t)}{dt^2} + \Omega_\theta^2 \delta \theta(t) = 0, \label{a16}
\end{eqnarray}
where the squared epicyclic frequencies are given by
\begin{eqnarray}
\Omega_r^2 = -\frac{1}{2 g_{rr} \dot{t}^2} \left. \frac{\partial^2 V_{\text{eff}}}{\partial r^2} \right|_{\theta = \frac{\pi}{2}}, \label{a17}\\
\Omega_\theta^2 = -\frac{1}{2 g_{\theta\theta} \dot{t}^2} \left. \frac{\partial^2 V_{\text{eff}}}{\partial \theta^2} \right|_{\theta = \frac{\pi}{2}}, \label{a18}
\end{eqnarray}
From Eqs.~(\ref{a17}) and (\ref{a18}), the radial and vertical epicyclic frequencies can be written as
\begin{eqnarray}
\nu_r &=& \frac{\Omega_r}{2\pi}, \\
\nu_\theta &=& \frac{\Omega_\theta}{2\pi}.
\end{eqnarray}
The explicit expressions of $\nu_\phi$, $\nu_r$, and $\nu_\theta$ for the dyonic KNKTN BH are given by Eqs.~(\ref{a12}), (\ref{a19}), and (\ref{a20}) in Appendix A. In the context of equatorial circular orbits for a test particle, the radial epicyclic frequency $\nu_{r}$ characterizes oscillations in the radial direction relative to the mean orbit, whereas the vertical epicyclic frequency $\nu_{\theta}$ describes oscillations perpendicular to the equatorial plane.

From the three fundamental frequencies discussed above, one can define the periastron precession frequency, $\nu_{\rm per}$, and the nodal precession frequency, $\nu_{\rm nod}$, as
\begin{eqnarray}
\nu_{\rm per} &=&  \nu_\phi - \nu_r, \\
\nu_{\rm nod} &=&  \nu_\phi - \nu_\theta.
\end{eqnarray}
Within the relativistic precession model for X-ray BH binaries~\cite{Stella:1997tc, Stella:1998mq, Stella:1999sj}, these three frequencies, $\nu_\phi$, $\nu_{\rm per}$, and $\nu_{\rm nod}$, are identified with the observed upper high-frequency QPO ($\nu_U$), lower high-frequency QPO ($\nu_L$), and low-frequency type-C QPO ($\nu_C$), respectively:
\begin{eqnarray}
\nu_U = \nu_\phi,\;\; \nu_L = \nu_{\rm per}, \;\;\nu_C = \nu_{\rm nod}.
\end{eqnarray}

In the dyonic KNKTN metric, all of these frequencies depend on the BH mass, spin, and the additional charges. The behavior of the fundamental QPO frequencies is illustrated in Figs.~\ref{1} and~\ref{2}. As shown in Fig.~\ref{1}, all three frequencies increase with increasing Taub-NUT charge. This trend can be attributed to the gravitomagnetic field introduced by the Taub-NUT charge, which twists the surrounding spacetime in a manner analogous to the frame-dragging effect produced by BH spin. The net result is a tightening of the circular orbits near the BH, leading to higher orbital velocities and consequently increased values of $\nu_{\phi}$, $\nu_{\rm per}$, and $\nu_{\rm nod}$. In contrast, Fig.~\ref{2} shows that all three frequencies decrease as the magnetic charge $P/M$ increases. This behavior arises because the magnetic charge acts to reduce the effective gravitational potential, shifting the circular orbits outward from the central object. As a result, the orbital velocity decreases, and both the radial and vertical epicyclic oscillations are suppressed, leading to lower QPO frequencies. Furthermore, in both figures, it is apparent that increasing the spin parameter $a/M$ leads to a decrease in all three frequencies, indicating that the combined effects of magnetic charge $P/M$ and Taub-NUT charge $n/M$ act to reduce the spin-induced frame-dragging.

\section{Constraints on extra charges of the dyonic KNKTN spacetime}
\renewcommand{\theequation}{3.\arabic{equation}} \setcounter{equation}{0}

In this section, we apply the relativistic precession model, together with the observed QPO frequencies from several BH X-ray binary systems, to constrain the parameters of the dyonic KNKTN spacetime. We focus on five well-studied QPO events, obtained from the extensively analyzed X-ray binary sources GRO J1655-40, XTE J1550-564, XTE J1859+226, GRS 1915+105, and H1743-322, as summarized in Table~\ref{tab: I}. To determine the allowed ranges of BH parameters, we perform MCMC analyses, presenting the resulting best-fit values and physically plausible parameter constraints.

\begin{table*}
\caption{The QPOs from the X-ray binaries that have been selected for investigation included their mass, orbital frequencies, periastron precession frequencies, and nodal precession frequencies.}
\label{tab: I}
\begin{ruledtabular}
\begin{tabular}{cccccc}
\; & GRO J1655-40  &XTE J1550-564  & XTE J1859+226 & GRS 1915+105  & H1743-322 
\\  
\hline 
$M~(M_{\odot})$ & $5.4 \pm 0.3$~\cite{Motta:2013wga} & $9.1 \pm 0.61$~\cite{Remillard:2002cy,Orosz:2011ki} & $7.85 \pm 0.46$~\cite{Motta:2022rku} & $12.4^{+2.0}_{-1.8}$~\cite{Remillard:2006fc} & $\gtrsim 9.29$~\cite{Ingram:2014ara} 
\\ \;\\
$\nu_{\phi}$ (Hz) &$441 \pm 2$~\cite{Motta:2013wga} & $276 \pm 3$~\cite{Remillard:2002cy} & $227.5^{+2.1}_{-2.4}$~\cite{Motta:2022rku} & $168 \pm 3$~\cite{Remillard:2006fc} & $240 \pm 3$~\cite{Ingram:2014ara}
\\ \;\\                  
$\nu_{\text{per}}$ (Hz) & $298 \pm 4$~\cite{Motta:2013wga} & $184 \pm 5$~\cite{Remillard:2002cy} & $128.6^{+1.6}_{-1.8}$~\cite{Motta:2022rku} & $113 \pm 5$~\cite{Remillard:2006fc} & $165^{+9}_{-5}$~\cite{Ingram:2014ara}  
\\\;\\
$\nu_{\text{nod}}$ (Hz) &  $17.3 \pm 0.1$~\cite{Motta:2013wga} & -- & $3.65 \pm 0.01$~\cite{Motta:2022rku} & -- & $9.44 \pm 0.02$~\cite{Ingram:2014ara}
\end{tabular}
\end{ruledtabular}
\end{table*}

\begin{table*}
\caption{We choose a Gaussian prior on the parameters \((M, a/M, r/M)\) of the dyonic KNKTN BH from QPOs for X-ray binaries and a uniform prior on $Q/M \in [0,3],\ P/M \in [0,3], \  n/M \in [0,1]$.}
\label{tab: II}
\begin{ruledtabular}
 \begin{tabular}{ccccccccccc} 
Parameters & \multicolumn{2}{c|}{GRO J1655--40} & \multicolumn{2}{c|}{XTE J1550--564} & \multicolumn{2}{c|}{XTE J1859+226} & \multicolumn{2}{c|}{GRS 1915+105} & \multicolumn{2}{c}{H1743--322} \\
 & $\mu$ & $\sigma$ & $\mu$ & $\sigma$ & $\mu$ & $\sigma$ & $\mu$ & $\sigma$ & $\mu$ & $\sigma$ \\
  \hline 
 $M~(M_{\odot})$ & 5.307 & 0.066 & 9.10 & 0.61 & 7.85 & 0.46 & 12.41 & 0.62 & 9.29 & 0.46 
 \\  \;\\
$a/M$   & 0.286 & 0.003 & 0.34 & 0.007 & 0.149 & 0.005 & 0.29 & 0.015 & 0.27 & 0.013 
\\\;\\
$r/M$   & 5.677 & 0.035 & 5.47 & 0.12 & 6.85 & 0.18 & 6.10 & 0.30 & 5.55 & 0.27 
\end{tabular}
\end{ruledtabular}
\end{table*}

\subsection{Markov chain Monte Carlo analysis}

In this study, we employ the MCMC technique implemented in the \emph{emcee} package~\cite{Foreman-Mackey:2012any} to constrain the parameters of the dyonic KNKTN BH. The posterior distribution is given by
\begin{equation}
\mathcal{P}(\Theta | \mathcal{D}, \mathcal{M}) = \frac{P(\mathcal{D} | \Theta, \mathcal{M}) \, \pi(\Theta | \mathcal{M})}{P(\mathcal{D} | \mathcal{M})},
\end{equation}
where $\pi(\Theta | \mathcal{M})$ and $P(\mathcal{D} | \Theta, \mathcal{M})$ denote the prior and likelihood, respectively. For the priors of parameters $(M, a/M, r/M)$, we adopt truncated Gaussian distributions of the form
\begin{equation}
\pi(\theta_i) \propto \exp\left[ -\frac{1}{2} \left( \frac{\theta_i - \theta_{0, i}}{\sigma_i} \right)^2 \right], \quad \theta_{\text{low}, i} < \theta_i < \theta_{\text{high}, i},
\end{equation}
where $\theta_{i} = [M, a/M, r/M]$ and $\sigma_i$ is the corresponding standard deviation and for extra charges, $(P/M, Q/M, n/M)$, we adopt the uniform prior in the analyses. The observational values and adopted priors for the dyonic KNKTN BH parameters are summarized in Tables~\ref{tab: I} and \ref{tab: II}.

Our MCMC analysis incorporates three distinct data sets based on the orbital, periastron precession, and nodal precession frequencies, derived as described in the previous section. The total likelihood function $\mathcal{L}$ is thus given by~\cite{Liu:2023vfh, Bambi:2013fea}
\begin{equation}
\log \mathcal{L}_{\rm tot} = \log \mathcal{L}_{\rm obs} + \log \mathcal{L}_{\rm per} + \log \mathcal{L}_{\rm nod},
\end{equation}
where 
\begin{equation}
\log \mathcal{L} = -\frac{1}{2} \sum_i \frac{\left(\Vec{D}_{\rm obs}^i - \Vec{D}_{\rm th}^i\right)^2}{(\sigma_i)^2}\,.
\end{equation}
Here, $\Vec{D}_{\rm obs}^i$ and $\Vec{D}_{\rm th}^i$ denote the $i$-th observed and theoretical data points, respectively, and $\sigma_i$ is the corresponding statistical uncertainty on the measurement. This form is adopted for each of the three frequencies, and the total log-likelihood is obtained by summing the individual contributions.

We constrain the dyonic KNKTN BH parameters across four distinct setups separately: 
\begin{enumerate}
    \item[(1)] $(M, a/M, r/M, Q/M)$ with magnetic charge $P = 0$ and Taub-NUT parameter $n = 0$;
    \item[(2)] $(M, a/M, r/M, P/M)$ with electric charge $Q = 0$ and $n = 0$;
    \item[(3)] $(M, a/M, r/M, n/M)$ with $Q = 0$ and $P = 0$;
    \item[(4)] $(M, a/M, r/M, Q/M, P/M, n/M)$.
\end{enumerate}
The results for these cases are presented in the following subsections. The prior range for all parameters in these setups is listed in Table~\ref{tab: II}.

\subsection{Constraints on electric charge}

In this subsection, we investigate a simplified scenario of the dyonic KNKTN BH by setting the magnetic charge $P = 0$ and the Taub-NUT parameter $n = 0$. Under these assumptions, the solution reduces to the electrically charged Kerr–Newman BH, which allows us to isolate and systematically study the influence of the electric charge $Q$ on the observed QPOs. To constrain the BH parameters, we perform MCMC analyses using QPO data from five stellar-mass X-ray binary systems: GRO J1655-40, XTE J1550-564, XTE J1859+226, GRS 1915+105, and H1743-322. In all cases, we find no significant evidence for a nonzero electric charge, indicating these BHs are essentially electrically neutral. Consequently, we can place stringent upper bounds on the ratio $Q/M$ for the BH in each source. The corresponding best-fit values of the mass $M$, dimensionless radius $r/M$, spin parameter $a/M$, and the upper limits on $Q/M$ for each system are reported in Table~\ref{tab: III}. Note that a combined constraint on $Q/M$ using three different QPO sources—GRO J1655-40, XTE J1859+226, and H1743-322—was derived in Ref.~\cite{Yang:2025aro} employing the $\chi^2$ method. Although a different treatment was adopted, the result is also consistent with the Kerr BH scenario.

\begin{figure*}
\centering
\includegraphics[scale=0.5]{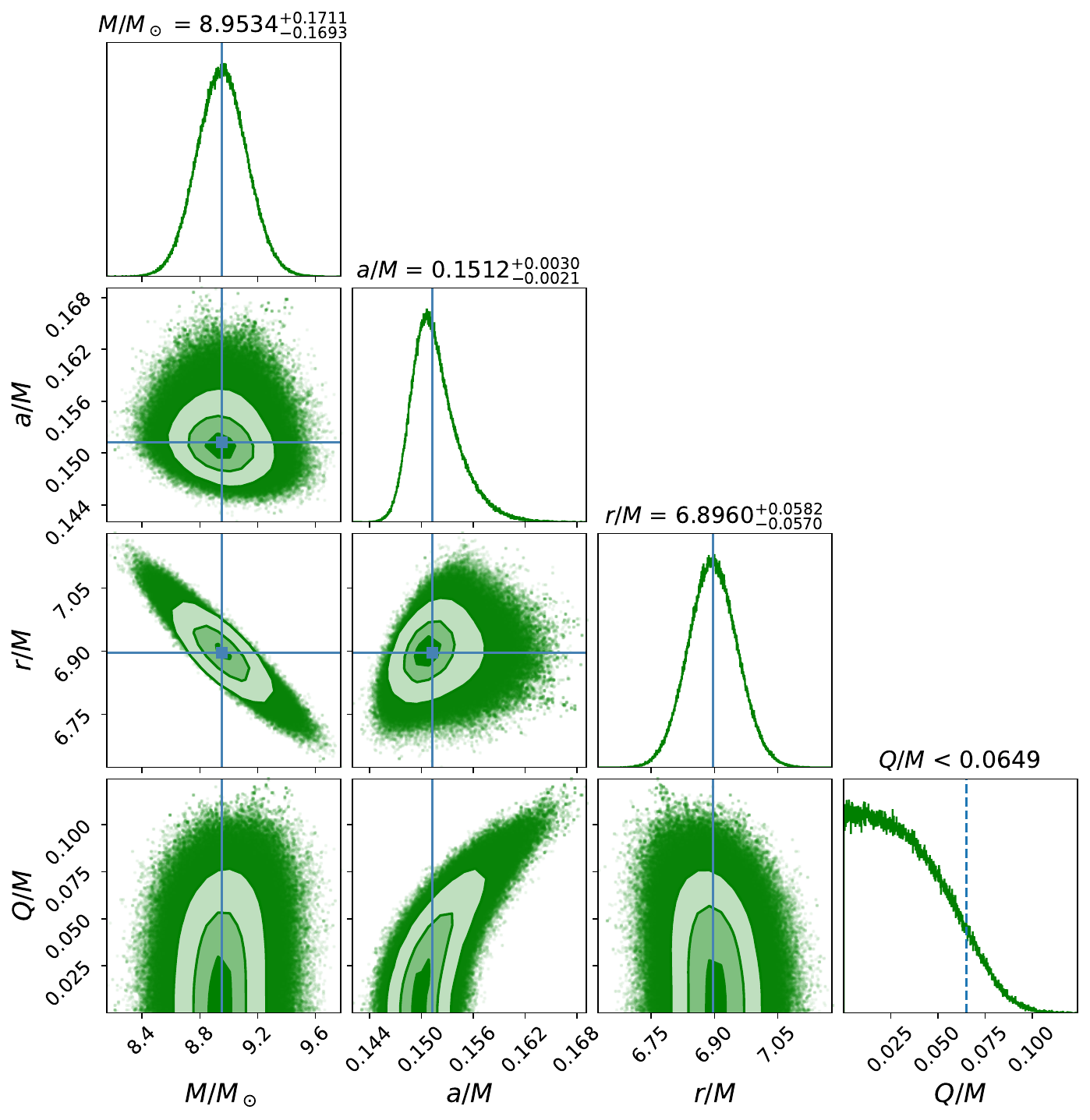}
\caption{Assuming $P/M=0$ and $n/M=0$, the dyonic KNKTN BH reduces to the electrically charged Kerr–Newman BH. The posterior distribution on the BH parameters is obtained from QPO observations of the XTE J1859+226 system.}
\label{1a}
\end{figure*}

Among the five sources analyzed, the tightest constraint on the dimensionless electric charge $Q/M$ arises from the system XTE J1550-564. Figure~\ref{1a} displays the posterior distributions for the physical parameters obtained from the MCMC analysis of this system. The diagonal panels show the marginalized one-dimensional probability distributions for individual parameters, while the off-diagonal panels illustrate joint probability contours that reveal correlations among pairs of parameters. The shaded regions correspond to the 68\% and 90\% confidence intervals, highlighting the ranges most strongly supported by the QPO data.

The QPO data from XTE J1859+226 provide an especially precise measurement of the BH's mass, with the posterior distribution sharply peaked around $8.95\,M_{\odot}$, consistent with the expected inverse scaling between fundamental QPO frequencies and BH mass. For this system, the dimensionless spin parameter $a/M$ is significantly lower than unity, indicating that the BH rotates far below the extremal (maximal spin) limit; this is reflected in its more modest frame-dragging effects, which modulate orbits near the innermost stable circular orbit (ISCO) and thus influence observable signals such as QPOs and gravitational waves. The inferred orbital radius $r/M$ is found to be close to the ISCO, with a small uncertainty, supporting the robustness of the model and the physical association of QPOs with nearly-ISCO orbits. From our analysis of XTE J1859+226, we derive a 90\% confidence upper limit on the electric charge:
\begin{eqnarray}
Q/M \leq 0.0649.
\end{eqnarray}
This stringent constraint implies that the BH is essentially electrically neutral, or that any electromagnetic influence on the QPO behavior is subdominant compared to gravitational effects.

\begin{table*}
\caption{The best-fit values of the electrically charged Kerr–Newman BH parameters from QPOs in X-ray binaries are reported at 68\% confidence intervals, while the parameter $Q/M$ is given at the 90\% confidence level.}
\label{tab: III}
\begin{ruledtabular}
\begin{tabular}{cccccc}
\; & GRO J1655--40  &XTE J1550--564  & XTE J1859+226 & GRS 1915+105 & H1743--322 \\ \hline 
$M/(M_{\odot})$ & $5.8369^{+0.0391}_{-0.0387}$  & $10.0287^{+0.2348}_{-0.2335}$  & $8.9534^{+0.1711}_{-0.1693}$ &$13.8663^{+0.4547}_{-0.4498}$ & $10.7982^{+0.2739}_{-0.2788}$ 
\\ \; \\
 $a/M$ &$0.2885^{+0.0020}_{-0.0020}$ & $0.3404^{+0.0069}_{-0.0069}$  & $0.1512^{+0.0030}_{-0.00321}$  & $0.2804^{+0.0146}_{-0.0146}$  &$0.2871^{+0.0070}_{-0.0070}$ 
 \\ \; \\
$r/M$ & $5.8027^{+0.0206}_{-0.0209}$ & $5.5862^{+0.0858}_{-0.0775}$  & $6.8960^{+0.0582}_{-0.0570}$& $6.1657^{+0.1315}_{-0.1269}$  & $5.7623^{+0.0846}_{-0.0852}$ 
\\  \; \\
$Q/M$ & $<0.1836$  &$<0.2592$  & $< 0.0649$  &$<0.2608$  &$< 0.3483$
\end{tabular}
\end{ruledtabular}
\end{table*}

\subsection{Constraints on magnetic charge}

This subsection examines the spatial configuration of the dyonic KNKTN BH assuming vanishing electric charge ($Q/M = 0$) and Taub-NUT parameter ($n/M = 0$). Our goal is to constrain the parameters ($M$, $r/M$, $a/M$, $P/M$) in this specific scenario. To achieve this, we perform MCMC analyses using the observed QPOs from five sources, as presented in Table~\ref{tab: I}. Similar to the results discussed in the previous subsection, we do not find any statistically significant evidence for a nonzero value of the magnetic charge for each analyzed source. This allows us to place stringent upper limits on $P/M$ for each BH. Table~\ref{tab: IV} summarizes the best-fit values of the parameters $(M, a/M, r/M)$, along with the corresponding upper limits on $P/M$ obtained from the five QPO sources.

As shown in Table~\ref{tab: IV}, the tightest constraint on the dimensionless magnetic charge $P/M$ is found for the GRO J1655-40 system. Figure~\ref{1a} displays the posterior distributions for the parameters $(M, r/M, a/M, P/M)$ based on the MCMC analysis of this X-ray binary. In this figure, the shaded triangular panels represent the correlations between pairs of parameters, while the diagonal panels show the one-dimensional marginalized posterior distributions for each parameter. The overlaid contours correspond to confidence levels of 68\% and 90\%, indicating the most probable regions for the true parameter values. For the GRO J1655-40 system, we derive a $90\%$ upper limit of
\begin{eqnarray}
    P/M \lesssim 0.1837  
\end{eqnarray}
These results indicate no observational evidence for a substantial magnetic charge in these BHs. Nonetheless, this constraint has important physical implications: in dyonic solutions, a nonzero magnetic charge $P/M$ alters the spacetime geometry and, in particular, influences the properties and locations of circular orbits around the BH.

\begin{table*}
\caption{The best-fit values of the magnetically charged Kerr BH parameters $M$,$a/M$, and $r/M$ from QPOs for the X-ray binaries at a 68\% confidence interval, while the upper bounds are mentioned for the parameter $P/M$ at a 90\% confidence level.}
\label{tab: IV}
\begin{ruledtabular}
\begin{tabular}{cccccc} 
\; & GRO J1655-40 \ &XTE J1550-564 \ & XTE J1859+226 \ & GRS 1915+105 \ & H1743-322
\\ \hline \\
 $M/(M_{\odot})$ & $5.8369^{+0.0392}_{-0.0391}$ & $10.1513^{+0.2416}_{-0.2349}$ & $8.9758^{+0.1781}_{-0.1732}$ &$13.8682^{+0.4555}_{-0.4488}$&$10.8006^{+0.2727}_{-0.2700}$ 
 \\ \; \\
$a/M$ &$0.2885^{+0.0020}_{-0.0020}$& $0.3392^{+0.0069}_{-0.0068}$ & $0.1487^{+0.0019}_{-0.0021}$ & $0.2805^{+0.0147}_{-0.0147}$ & $0.2871^{+0.0071}_{-0.0071}$ 
\\ \; \\
 $r/M$ & $5.8027^{+0.0206}_{-0.0209}$& $5.5120^{+0.0778}_{-0.0816}$ & $6.8766^{+0.0638}_{-0.0685}$& $6.1644^{+0.1319}_{-0.1269}$ & $5.7618^{+0.0850}_{-0.0854}$ 
 \\ \; \\
$P/M$ & $< 0.1837$ &$<0.3092$  & $< 0.2592$  & $<0.2618$  &$<0.3484$
 \end{tabular}
\end{ruledtabular}
\end{table*}

\begin{figure*}
\centering
\includegraphics[scale=0.5]{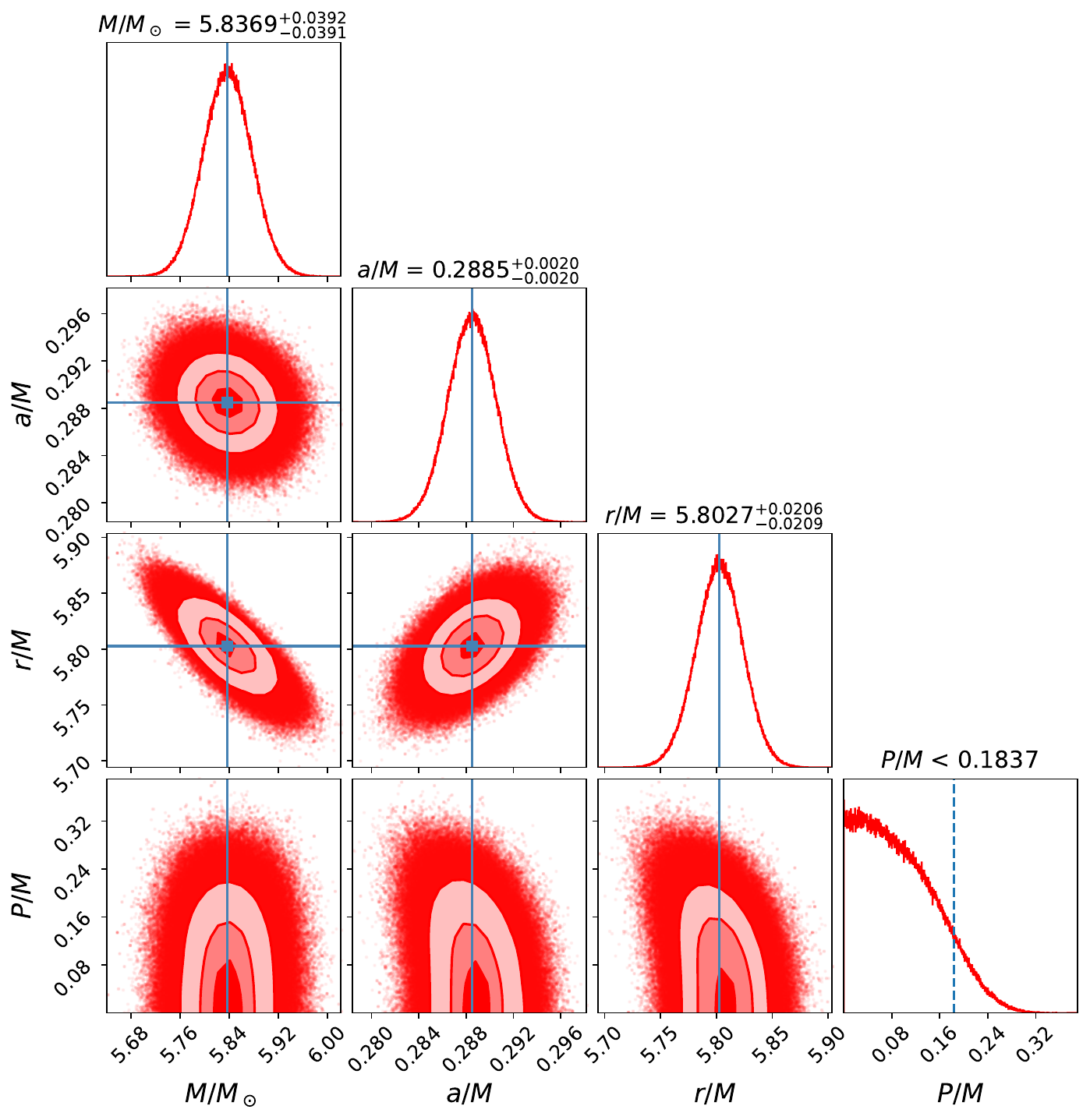}
\caption{By setting $Q/M=0$ and $n/M=0$, the dyonic KNKTN BH reduces to the magnetically charged Kerr BH. We constrain its parameters using observational data of QPOs from the GRO J1655-40 QPO system. }\label{4}
\end{figure*}

\subsection{Constraints on Taub-NUT parameter}

In this subsection, we focus on the spatial case of the dyonic KNKTN BH with vanishing electric ($Q/M = 0$) and magnetic ($P/M = 0$) charges. Under these constraints, the dyonic KNKTN solution reduces to the well-known Kerr–Taub–NUT BH. Our primary objective is to constrain the parameters $(M, a/M, r/M, n/M)$ of this particular solution using observational QPO data obtained from five X-ray binary systems. To this end, we employ MCMC analysis utilizing the QPO data summarized in Table~\ref{tab: I}. The resulting best-fit values for the parameters $(M, a/M, r/M)$ are reported in Table~\ref{tab: V}.

From our analysis, we find no compelling evidence for a nonzero Taub–NUT parameter in four of the systems—GRO J1655–40, XTE J1859+226, XTE J1550–564, and H1743–322. However, for the source GRS 1915+105, the results indicate a statistically significant nonzero value of the Taub–NUT parameter, implying a nonzero value for the gravitomagnetic monopole moment. 

\begin{table*}
 \caption{Best-fit values for the Kerr–Taub–NUT BH parameters obtained from QPO data for the X-ray binaries, quoted at 68\% confidence interval. Upper bounds for the parameter $n/M$ are given at the 90\% confidence level.}
 \label{tab: V}
\begin{ruledtabular}
\begin{tabular}{cccccc}
\; & GRO J1655-40 \ &XTE J1550--564 \ & XTE J1859+226 \ & GRS 1915+105 \ & H1743--322 
\\ \hline 
$M/(M_{\odot})$ & $5.8377^{+0.0392}_{-0.0389}$ & $10.0305^{+0.2360}_{-0.2339}$ & $8.9513^{+0.1717}_{-0.1690}$&$13.2969^{+0.4833}_{-0.4716}$&$10.7861^{+0.2676}_{-0.2651}$
\\ \; \\
$a/M$ &$0.2894^{+0.0021}_{-0.0020}$& $0.3405^{+0.0069}_{-0.0069}$ & $0.1512^{+0.0030}_{-0.0021}$ & $0.2889^{+0.0148}_{-0.0148}$ & $0.2915^{+0.0072}_{-0.0067}$ 
\\ \; \\
$r/M$ &$5.8054^{+0.0200}_{-0.0199}$& $5.5856^{+0.0857}_{-0.0778}$& $6.8966^{+0.0582}_{-0.0572}$&  $6.5716^{+0.2034}_{-0.2006}$& $5.7797^{+0.0799}_{-0.0762}$  
\\ \;\\
 $n/M$ & $< 0.0446$   & $<0.2574$ & $< 0.0647$  & $0.5435^{+0.1031}_{-0.1256}$  & $<0.0815$
\end{tabular}
\end{ruledtabular}
\end{table*}

\begin{figure*}
\centering
\includegraphics[scale=0.27]{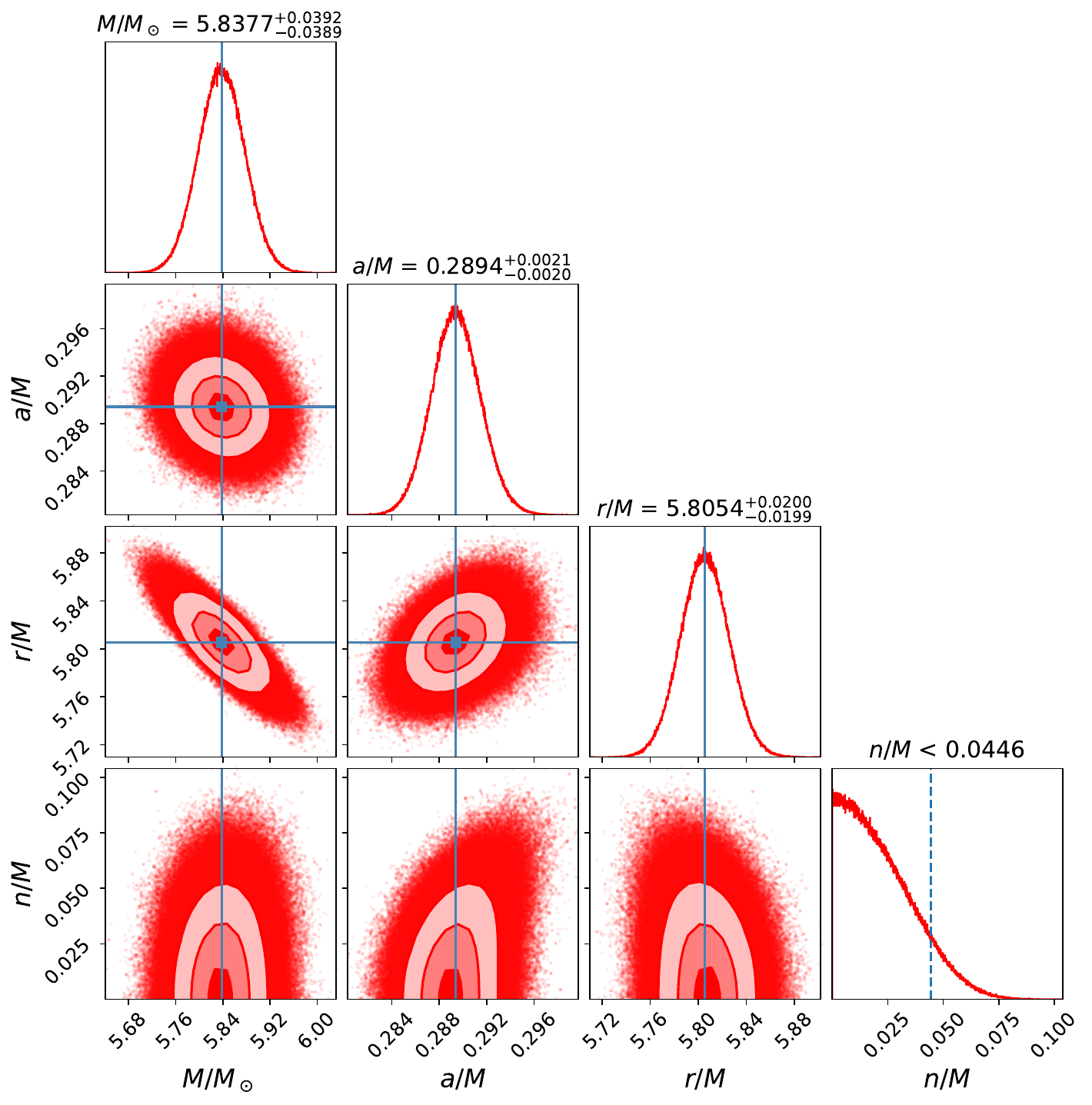}
\includegraphics[scale=0.27]{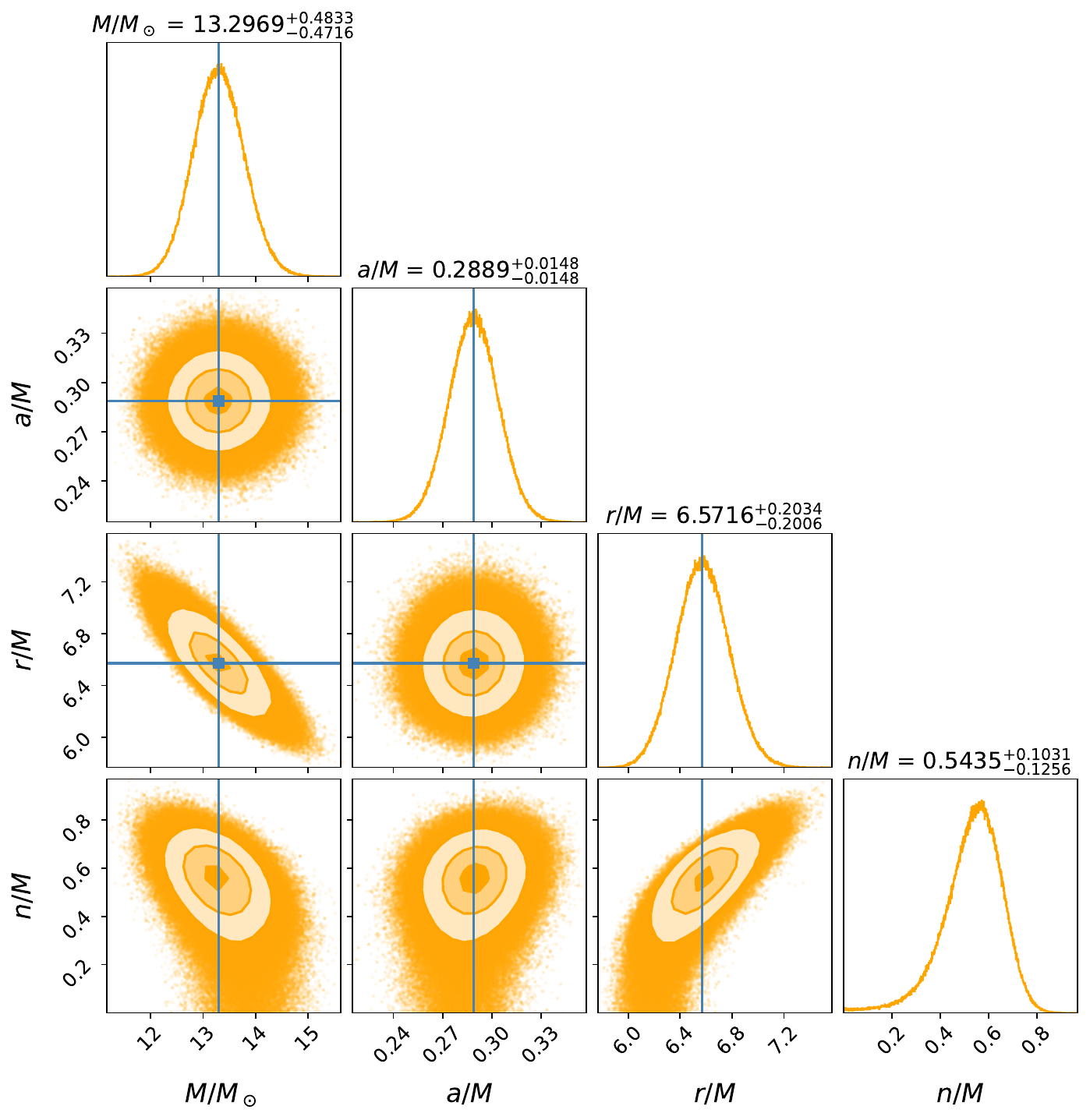}
\caption{Constraints on the Kerr–Taub–NUT BH parameters derived from QPO data of GRO J1655–40 (red contours) and GRS 1915+105 (orange contours), obtained after setting $Q/M = 0$ and $P/M = 0$.}
\label{5}
\end{figure*}

Among the four sources with no evidence for a Taub-NUT charge, GRO J1655–40 provides the most stringent upper limit on the dimensionless NUT parameter $n/M$. The results of the MCMC analysis for this source are summarized in Fig.~\ref{5}. The diagonal panels of Fig.~\ref{5} show the one-dimensional marginalized distributions for each parameter, while the off-diagonal panels illustrate the correlations between parameter pairs. Confidence contours for the joint posterior distributions are also displayed, corresponding to 68\% and 90\% confidence levels. Based on this analysis, we find that
\begin{eqnarray}
    n/M \lesssim 0.0446
\end{eqnarray}
at the 90\% confidence level, implying that BH in GRO J1655–40 does not harbor a significant Taub-NUT charge. In addition, Ref.~\cite{Yang:2025aro} employed the $\chi^2$ method to derive a combined constraint on $n/M$ using three distinct QPO sources—GRO J1655-40, XTE J1859+226, and H1743-322—and found a potential result for a non-zero Taub-NUT parameter. This result is different from ours, but it should be noted that their analysis adopts a fundamentally different methodology.

In contrast, for GRS 1915+105, our analysis reveals a clear indication for a nonzero Taub–NUT parameter:
\begin{eqnarray}
    \frac{n}{M} = 0.5435^{+0.1031}_{-0.1256}
\end{eqnarray}
at 68\% confidence. The posterior distributions for the parameters $(M, a/M, r/M, n/M)$ of GRS 1915+105 are also presented in Fig.~\ref{5}. It is evident that the distribution of $n/M$ peaks well away from zero, suggesting a possible deviation from the pure Kerr geometry in this BH candidate.

\subsection{Posterior Constraints on all parameters for the dyonic KNKTN BH}

\begin{figure*}
\centering
\includegraphics[scale= 0.5]{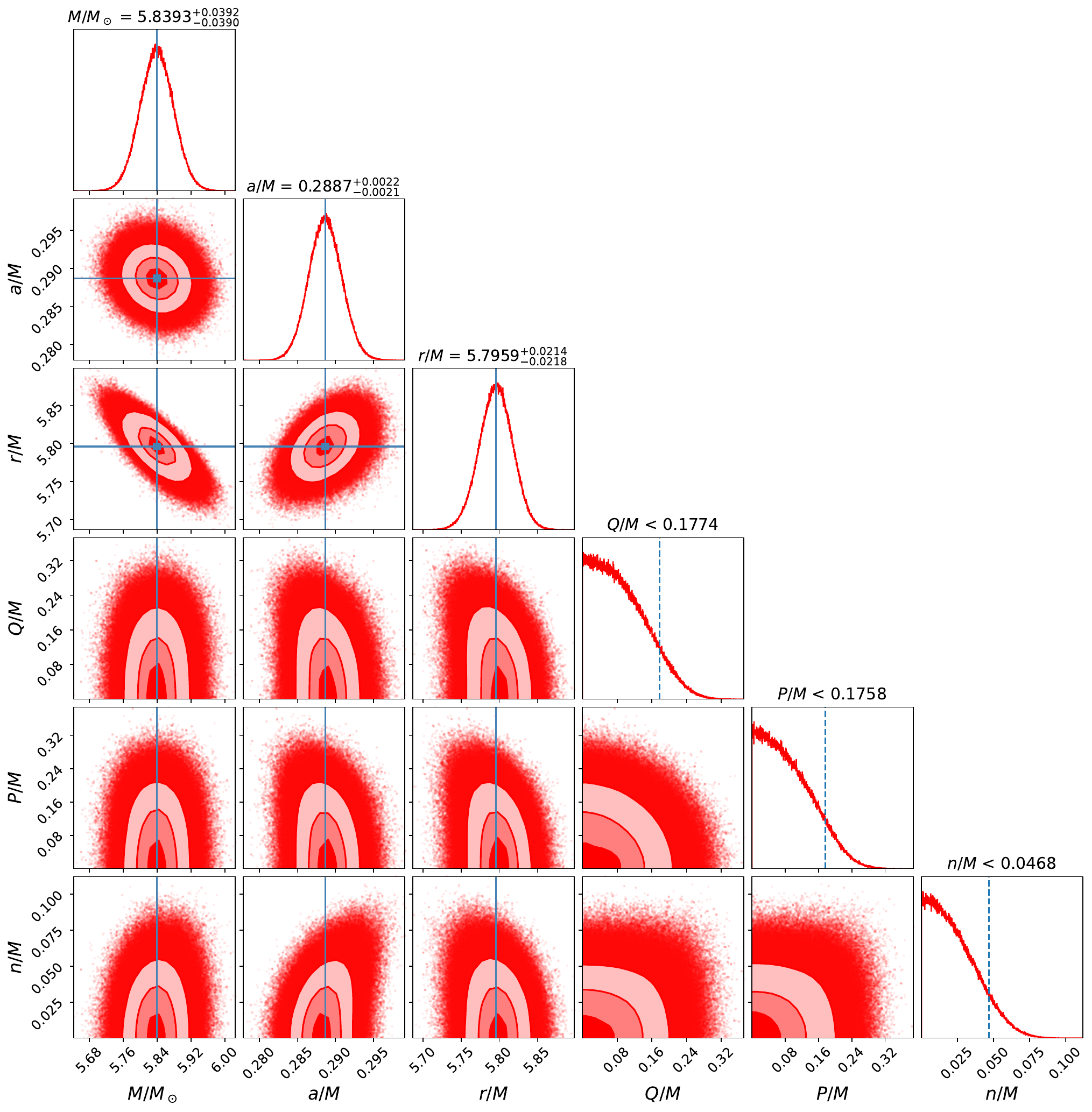}
\caption{1D, 2D marginallized distribution of model parameters  of the dyonic KNKTN BH with GRO J1655-40 current observations of QPOs within the relativistic
precession model.}\label{7}
\end{figure*}

\begin{figure*}
\centering
\includegraphics[scale=0.26]{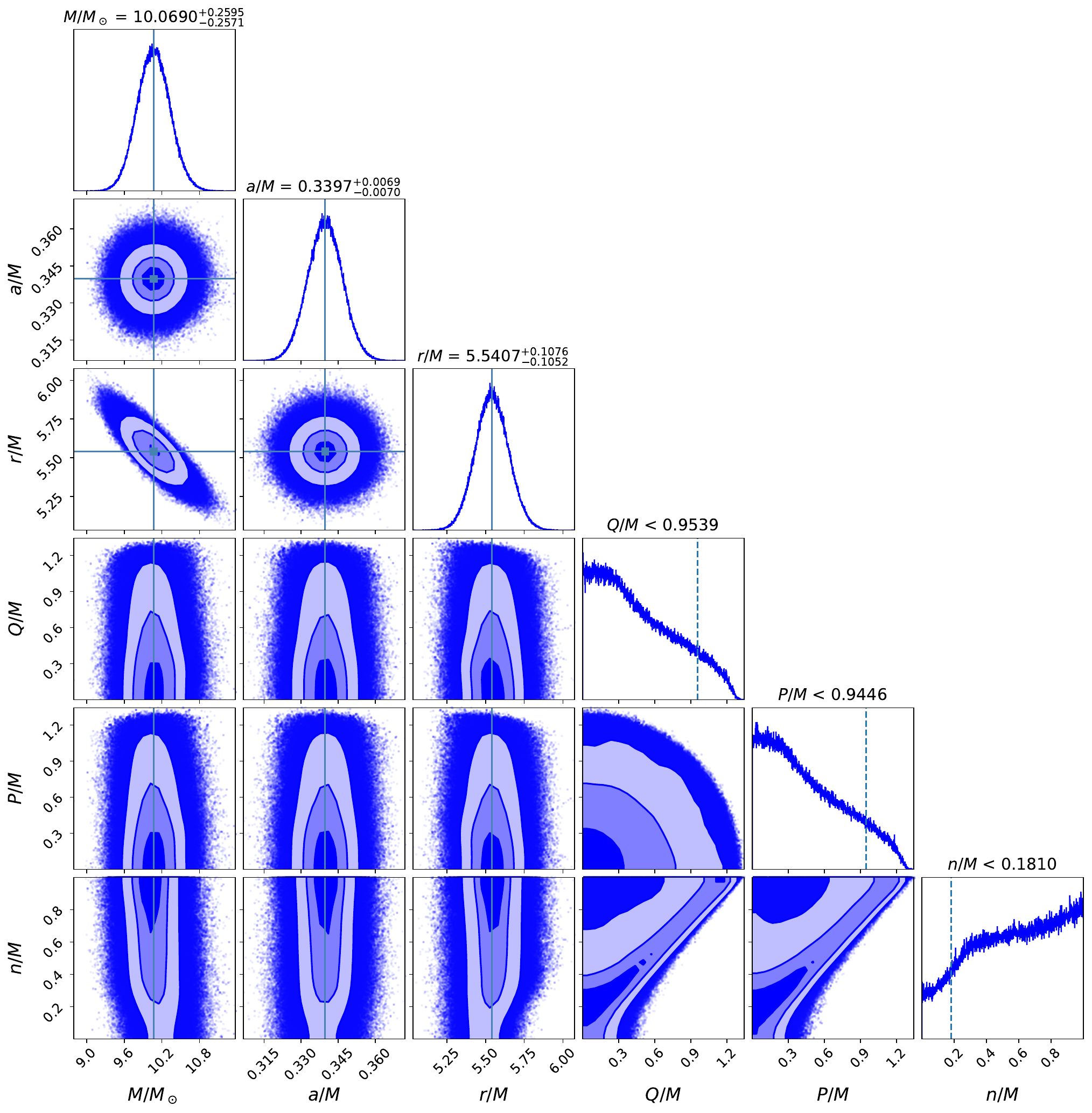}
\includegraphics[scale= 0.26]{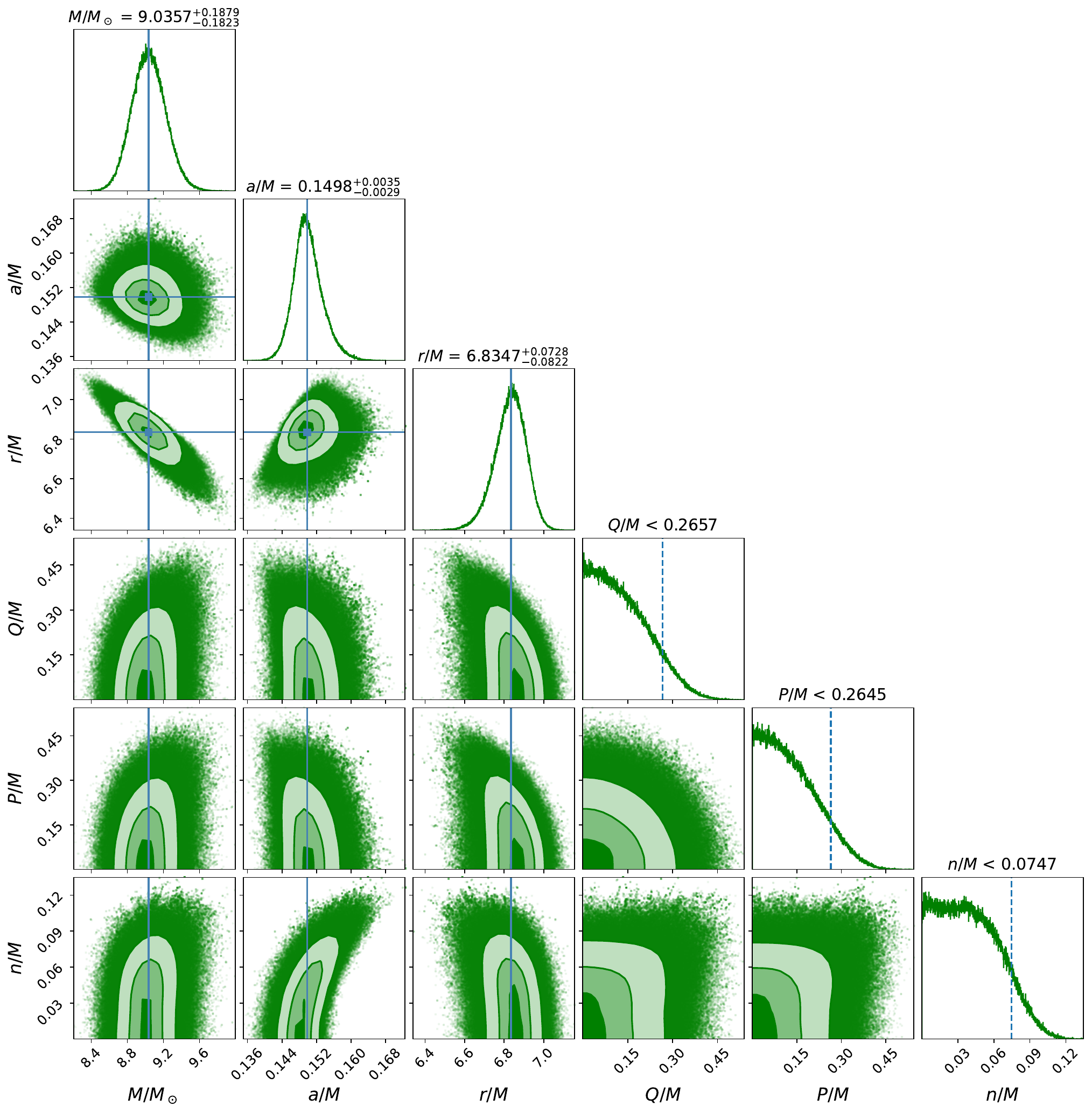}
\includegraphics[scale=0.26]{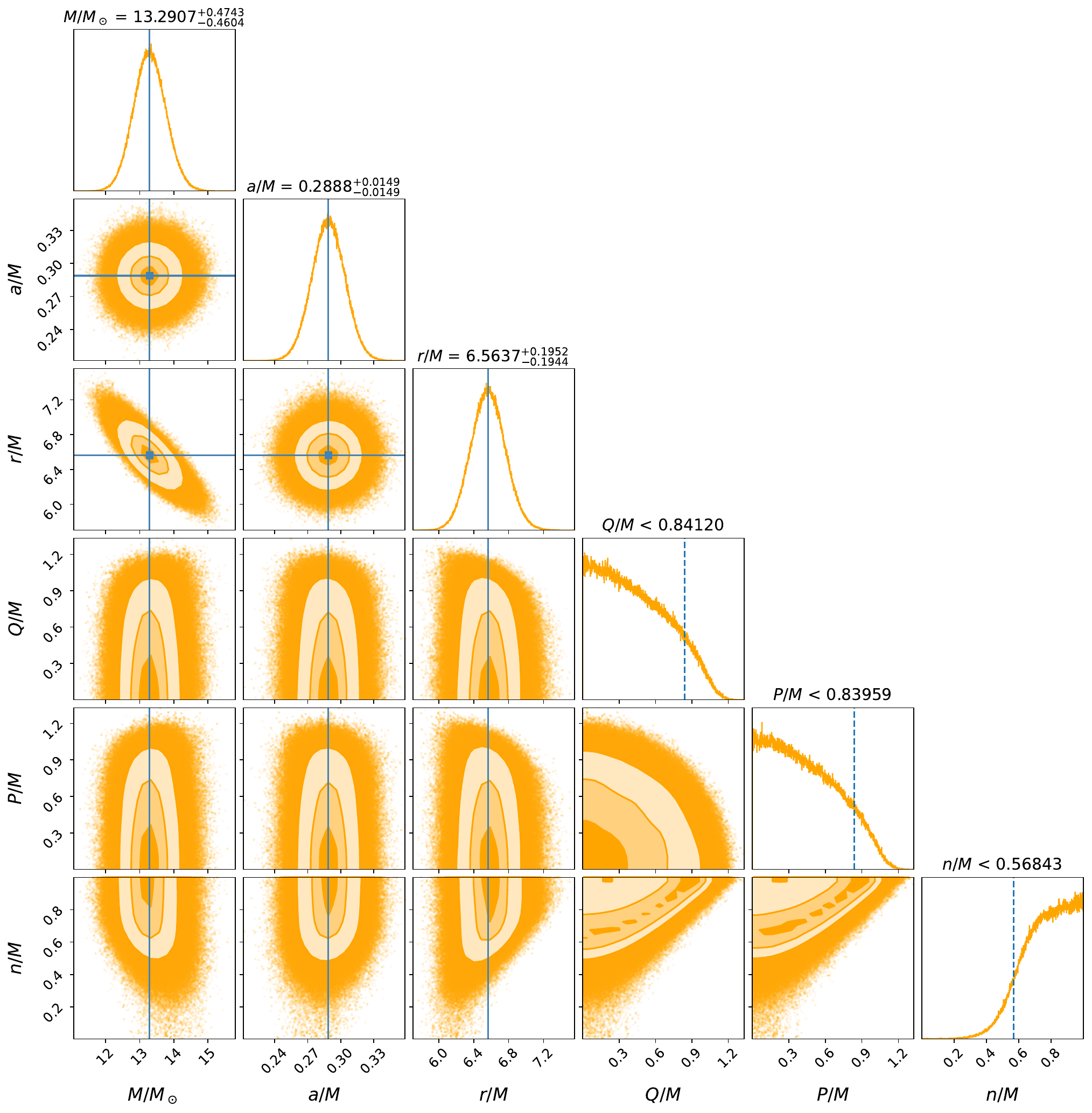}
\includegraphics[scale= 0.26]{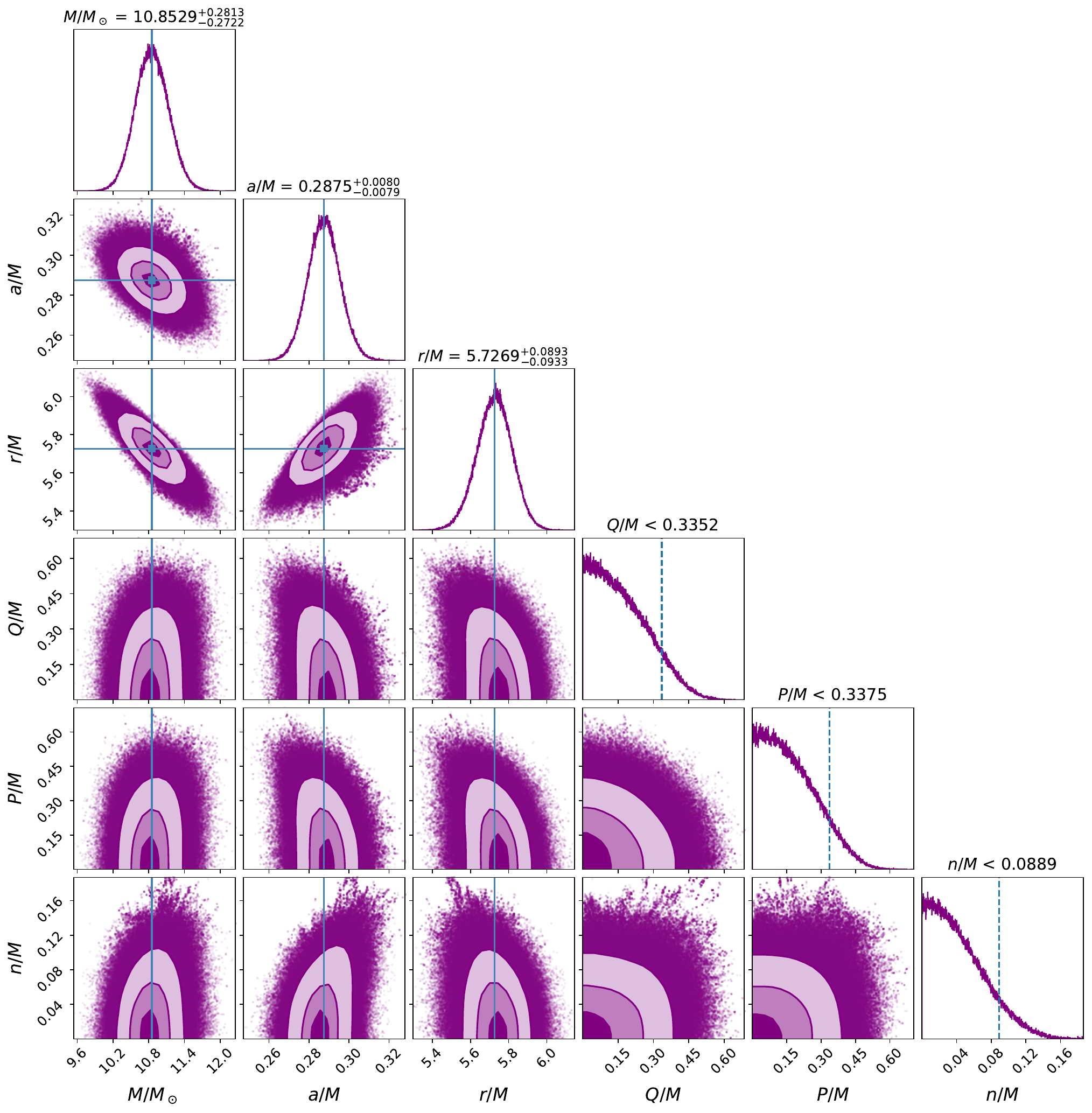}
\caption{1D, 2D marginalized distribution of model parameters of the dyonic KNKTN BH with XTE XTE J1550–564 (blue contours), XTE J1859+226 (green contours), GRS 1915+105 (orange contours), and H1743–322 (purple contours) from current observations of QPOs within the relativistic precession model.}\label{8}
\end{figure*}

This study explores the 6-dimensional parameter space $(M, a/M, r/M, Q/M, P/M, n/M)$ of the dyonic KNKTN BH using MCMC analysis. Figs.~\ref{7} and~\ref{8} present the full posterior distributions for all six parameters within the relativistic precession model. These figures summarize the MCMC results for the dyonic KNKTN BH as applied to the X-ray binaries GRO J1655–40, XTE J1550–564, XTE J1859+226, GRS 1915+105, and H1743–322, respectively. In the contour plots, the shaded regions indicate the 68\% and 90\% confidence intervals for each pair of parameters. The corresponding best-fit values and constraints for all parameters are provided in Table~\ref{tab: VI}.

\begin{table*}
 \caption{The best-fit values of the dyonic KNKTN BH parameters $M$, $a/M$, and $r/M$ are reported with 68\% confidence intervals based on QPO data from X-ray binaries, while the parameters $Q/M$, $P/M$, and $n/M$ are constrained at the 90\% confidence level.}
\label{tab: VI}
\begin{ruledtabular}
\begin{tabular}{cccccc}\
\; & GRO J1655-40 \ &XTE J1550-564 \ & XTE J1859+226 \ & GRS 1915+105 \ & H1743-322
\\ \hline  \\
$M/(M_{\odot})$ & $5.8393^{+0.0392}_{-0.0390}$ & $10.0690^{+0.2595}_{-0.2571}$ & $9.0357^{+0.1879}_{-0.1823}$ & $13.2907^{+0.4743}_{-0.4604}$& $10.8529^{+0.2813}_{-0.2722}$ 
\\ \;  \\
 $a/M$ &$0.2887^{+0.0022}_{-0.0021}$& $0.3397^{+0.0069}_{-0.0070}$ & $0.1498^{+0.0035}_{-0.0029}$ & $0.2888^{+0.0149}_{-0.0149}$ & $0.2875^{+0.0080}_{-0.0079}$ 
 \\ \; \\
 $r/M$ & $5.7959^{+0.0214}_{-0.0218}$ & $5.5407^{+0.1076}_{-0.1052}$ & $6.8347^{+0.0728}_{-0.0822}$ & $6.5637^{+0.1952}_{-0.1944}$ & $5.7269^{+0.0893}_{-0.0933}$
 \\\;\\
 $Q/M$ &  $<0.1774$   &$<0.9539$  & $<0.2657$ & $< 0.84120$ & $<0.3352$
 \\\;\\
$P/M$ & $<0.1758$  & $<0.9446$  & $< 0.2645$  & $<0.83959$  & $<0.3375$ 
\\\;\\
 $n/M$ & $<0.0468$  & not-constrained    & $<0.0747$   & $>0.568434$   &  $<0.0889$
\end{tabular}
 \end{ruledtabular}
\end{table*}

From Table~\ref{tab: VI}, we observe that the orbital radii for the five X-ray binaries range from $5.5856$ to $6.8966$ (in units of $r/M$). These radii lie close to the BH event horizon, where the gravitational field is particularly strong. In this region, physical effects such as frame-dragging due to BH spin, as well as the influence of electric charge, magnetic charge, and the Taub-NUT parameter, become significant and strongly impact particle orbits. For example, the orbital radius value $r/M=6.8966$ from XTE J1859+226 suggests that QPOs in this system may originate farther from the BH horizon, while the smaller value $r/M=5.5856$ for XTE J1550-564 points to oscillations occurring closer to the event horizon. Our analysis shows that the orbital radius $r/M=5.5856$ and its corresponding observed frequencies for XTE J1550-564 are highly sensitive to the BH mass, spin, electric and magnetic charges, and the Taub-NUT parameters of the dyonic KNTN BH, as particles in this region are subject to intense gravitational and relativistic effects.

To assess the validity of the dyonic KNKTN spacetime, we compare the model-predicted QPO frequencies at the best-fit parameter values with the corresponding observational data, as shown in Fig.~\ref{8}. Our results do not reveal statistically significant evidence supporting the existence of a dyonic KNKTN spacetime in the considered systems. Accordingly, we provide upper limits for the dimensionless parameters $Q/M$, $P/M$, and $n/M$ based on the GRO J1655–40 model, with constraints $Q/M<0.1774$, $P/M<0.1758$, and $n/M<0.0468$ at the $90\%$ confidence level. Notably, for the system GRS 1915+105, we infer a lower bound for the dimensionless Taub-NUT parameter, $n/M>0.568434$, at the $90\%$ confidence level, indicating a statistically significant presence of a non-vanishing Taub-NUT parameter. This result suggests the existence of strong gravitomagnetic effects, independent of classical spin, which distinguish dyonic KNKTN spacetimes from traditional Kerr BHs where frame-dragging arises solely from rotation.

Finally, our results highlight that the most stringent constraints are obtained for GRO J1655-40, which offers a comprehensive set of QPO observations, including orbital, periastron precession, and nodal precession frequencies, all measured with small statistical uncertainties. This rich dataset fulfills the requirements of the relativistic precession model and enables robust parameter estimation within the dyonic KNKTN spacetime. Consequently, GRO J1655-40 is particularly suitable for assessing possible deviations from the standard Kerr paradigm. The results presented in Figs.~\ref{1a} to \ref{8} and Tables~\ref{tab: III} to \ref{tab: VI} demonstrate full compatibility of these systems with a central BH described by the Kerr spacetime in general relativity.

\section{Conclusions}
\renewcommand{\theequation}{4.\arabic{equation}} \setcounter{equation}{0}

In this article, we have analyzed the QPOs observed in X-ray binary systems by investigating the epicyclic motion of particles near a dyonic KNKTN BH. We derived fundamental expressions for the epicyclic frequencies of particles in equatorial circular orbits, focusing on how the dimensionless electric charge ($Q/M$), magnetic charge ($P/M$), and Taub-NUT parameter ($n/M$) influence these frequencies. Our results show that increasing the Taub–NUT parameter $n/M$ enhances the gravitomagnetic effects, strengthening the effective gravitational attraction and increasing the fundamental frequencies. Conversely, increasing the magnetic charge $P/M$ leads to a decrease in the frequencies due to the introduction of a repulsive component in the gravitational potential from the magnetic field. We have also demonstrated that the parameters of the dyonic KNKTN BH significantly affect both the radial effective potential and the angular momentum of orbiting particles, thereby impacting the characteristic frequencies of their epicyclic oscillations—features that are closely linked to the QPO phenomena reported in astrophysical observations.

To test these theoretical predictions, we confronted our model with observational QPO data from five X-ray binaries—GRO J1655–40, XTE J1550–564, XTE J1859+226, GRS 1915+105, and H1743–322—using an MCMC approach to constrain the physical parameters of the central compact objects. Our joint analysis places stringent upper limits on the electric charge ($Q/M$), magnetic charge ($P/M$), and Taub–NUT parameter ($n/M$), with no significant evidence for nonzero values of $Q/M$ or $P/M$ across all sources. Notably, no substantial deviations from the classical Kerr metric—in particular, no significant Taub–NUT contribution—were found for GRO J1655–40, XTE J1859+226, XTE J1550–564, and H1743–322.

However, the QPO data from GRS 1915+105 present an intriguing exception. Our MCMC results for this source suggest a preference for a nonzero Taub–NUT parameter, hinting at an appreciable gravitomagnetic monopole moment and thus a possible deviation from the standard Kerr picture. This finding implies that, for GRS 1915+105, the BH might possess an intrinsic spacetime twist independent of angular momentum, as allowed in the dyonic KNKTN geometry, offering a novel mechanism for QPO generation beyond the classical frame-dragging effect.

In summary, while our constraints largely favor the conventional Kerr BH solution for most sources, the potential indication of a nonzero Taub–NUT parameter in GRS 1915+105 motivates further observational and theoretical investigation. Future high-precision QPO measurements will be essential to confirm or refute these results, deepening our understanding of strong-field gravity and the true nature of astrophysical BHs.

\renewcommand{\theequation}{5.\arabic{equation}} \setcounter{equation}{0}

\section*{Acknowledgements}

We would like to thank Cheng Liu and Xu-Jie Zhu for valuable discussions. This work is supported by the National Natural Science Foundation of China under Grants No.~12275238, the Zhejiang Provincial Natural Science Foundation of China under Grants No.~LR21A050001 and No.~LY20A050002, the National Key Research and Development Program of China under Grant No. 2020YFC2201503, and the Fundamental Research Funds for the Provincial Universities of Zhejiang in China under Grant No.~RF-A2019015. 

\appendix

\section{Appendix A: The expressions of three fundamental frequencies}
\renewcommand{\theequation}{A.\arabic{equation}} \setcounter{equation}{0}

For a dyonic KNKTN BH, the three fundamental frequencies $\nu_\phi$, $\nu_r$, and $\nu_\theta$ are given by
\begin{widetext}
\begin{eqnarray}
v_{\phi}=\frac{\Omega_{\phi}}{2\pi}&=&
\frac{ M a{_*}(M (n-r) (n+r)+r(-2 n^2+P^2+Q^2))}{M^2 a{_*}^2(M (n-r) (n+r)+r)+r(n^2+r^2)^2}+\frac{(n^2+r^2)^2\sqrt{-\frac{r (M (n-r) (n+r)+r(-2 n^2+P^2+Q^2))}{(n^2+r^2)^2}}}{M^2 a{_*}^2(M (n-r) (n+r)+r)+r(n^2+r^2)^2},\label{a12}
\end{eqnarray}
\begin{eqnarray}
\nu_{r} &= &\nu_{\phi}\Bigg[\frac{4 r^2 \left(2 r (r-M)-M a_*^2+P^2+Q^2\right)}{\left(n^2+r^2\right)^2}+\frac{8M a_* r \sqrt{-\frac{r(M (n-r) (n+r)+r(-2 n^2+P^2+Q^2))}{(n^2+r^2)^2}}}{n^2+r^2}+\frac{4 r (M-2 r)}{n^2+r^2}+\frac{M}{M-2 r}+ \nonumber \\&& \frac{M(M a_*^2 (M-2 r)-(M-r)(2 r (M-r)-P^2-Q^2))}{(M-2 r)(M \left(r^2-n^2\right)+r(2 n^2-P^2-Q^2))}\Bigg]^{1/2},\label{a19}
\end{eqnarray}
and 
\begin{eqnarray}
\nu_{\theta} &=& \nu_{\phi}\Bigg[ -\Bigg(\frac{2 n^2 r^3(8 M^2-2 n^2+3(P^2+Q^2))+ M a_*^2(-M (n^4+6 n^2 r^2-3 r^4)+2 n^2 r(P^2+Q^2+4 r^2)-2 r^3(P^2+Q^2))-M n^6}{(n^2+r^2)^2 (M (n-r) (n+r)+r (-2 n^2+P^2+Q^2))}-\nonumber\\&& \frac{2M a_*(n^2+r^2)(2 M r+2 n^2-P^2-Q^2)\sqrt{-r(M (n-r) (n+r)+r(-2 n^2+P^2+Q^2))}+M n^2 r^2(15 n^2-16(P^2+Q^2))}{(n^2+r^2)^2 (M (n-r) (n+r)+r (-2 n^2+P^2+Q^2))}-\nonumber\\&& \frac{15 M n^2 r^4 +M r^6+r^5(6 n^2- P^2-Q^2)+n^2 r(6 n^4-9 n^2(P^2+Q^2)+4(P^2+Q^2)^2)}{(n^2+r^2)^2 (M (n-r) (n+r)+r (-2 n^2+P^2+Q^2))}\Bigg)\Bigg]^{1/2},\label{a20}
\end{eqnarray} 
where $a_*\equiv a/J$.
\end{widetext}

\end{document}